\documentclass[aps,pra,reprint,superscriptaddress]{revtex4-2}
\usepackage{amsfonts,amssymb,amsthm,bm,mathrsfs,mathtools}
\usepackage[dvipsnames]{xcolor}
\usepackage{textgreek}
\usepackage[unicode]{hyperref}
\hypersetup{colorlinks=true,allcolors=blue,citecolor=magenta}
\usepackage{graphicx}
\usepackage{multirow}

\newcommand{\cE}{\mathcal{E}}

\newcommand{\cD}{\mathcal{D}}
\newcommand{\cI}{\mathcal{I}}
\newcommand{\cN}{\mathcal{N}}

\newcommand{\cP}{\mathcal{P}}
\newcommand{\cU}{\mathcal{U}}

\newcommand{\rS}{\mathrm{S}}
\newcommand{\rB}{\mathrm{B}}
\newcommand{\rSB}{\mathrm{SB}}

\DeclareMathOperator{\tr}{tr}
\DeclareMathOperator{\trB}{tr_B}
\DeclareMathOperator{\trS}{tr_S}

\DeclareMathOperator{\infid}{InF}
\newcommand{\transpose}{{\mkern-1.5mu\mathsf{T}}}

\newcommand{\upi}{\mathrm{i}}
\newcommand{\upe}{\mathrm{e}}
\newcommand{\upd}{\mathrm{d}}

\newcommand{\up}[1]{^{(#1)}}

\newcommand{\ad}{\mathrm{ad}}

\newcommand{\HB}{H_\mathrm{B}}
\newcommand{\HSB}{H_\mathrm{SB}}
\newcommand{\Heff}{H_\mathrm{eff}}

\newcommand{\rhoS}{\rho_\mathrm{S}}
\newcommand{\rhoB}{\rho_\mathrm{B}}

\newcommand{\phiB}{\phi_\rB}
\newcommand{\phiSB}{\phi_\rSB}

\newcommand{\Odd}{\Omega_{\DD}}
\newcommand{\DD}{\mathsf{DD}}
\newcommand{\PDD}{\mathsf{PDD}}
\newcommand{\CDD}{\mathsf{CDD}}
\newcommand{\CDDn}{\mathsf{CDD} n}

\newcommand{\ep}{\Phi_\mathrm{SB}}

\newcommand{\opnorm}[1]{\Vert\hspace*{-0.3mm}|#1|\hspace*{-0.3mm}\Vert}
\newcommand{\Btrinorm}{\Big\Vert\hspace*{-0.3mm}\Big|}

\newcommand{\norm}[2][]{#1\Vert {#2} #1\Vert}
\newcommand{\Ohat}{\widehat{\Omega}}
\newcommand{\wt}[1]{\widehat{#1}}

\newcommand{\Eqref}[1]{Eq.~\!(\ref{#1})}

\begin{document}

\title{Efficacy of noisy dynamical decoupling}
\author{Jiaan Qi}
\affiliation{Beijing Academy of Quantum Information Sciences, China}
\affiliation{Yale-NUS College, Singapore}

\author{Xiansong Xu}
\affiliation{Science, Mathematics and Technology Cluster, Singapore University of Technology and Design, 8 Somapah Road, 487372 Singapore}

\author{Dario Poletti}
\affiliation{Science, Mathematics and Technology Cluster, Singapore University of Technology and Design, 8 Somapah Road, 487372 Singapore}
\affiliation{EPD Pillar, Singapore University of Technology and Design, 8 Somapah Road, 487372 Singapore} 
\affiliation{Centre for Quantum Technologies, National University of Singapore} 
\affiliation{The Abdus Salam International Centre for Theoretical Physics, Strada Costiera 11, 34151 Trieste, Italy}
\affiliation{MajuLab, International Joint Research Unit UMI 3654,
CNRS-UCA-SU-NUS-NTU, Singapore}

\author{Hui Khoon Ng}
\affiliation{Yale-NUS College, Singapore}
\affiliation{Centre for Quantum Technologies, National University of Singapore}
\affiliation{Department of Physics, National University of Singapore}
\affiliation{MajuLab, International Joint Research Unit UMI 3654,
CNRS-UCA-SU-NUS-NTU, Singapore}

\begin{abstract}
Dynamical decoupling (DD) refers to a well-established family of methods for error mitigation, comprising pulse sequences aimed at averaging away slowly evolving noise in quantum systems. Here, we revisit the question of its efficacy in the presence of noisy pulses in scenarios important for quantum devices today: pulses with gate control errors, and the computational setting where DD is used to reduce noise in every computational gate. We focus on the well-known schemes of periodic (or universal) DD, and its extension, concatenated DD, for scaling up its power. The qualitative conclusions from our analysis of these two schemes nevertheless apply to other DD approaches. In the presence of noisy pulses, DD does not always mitigate errors. It does so only when the added noise from the imperfect DD pulses do not outweigh the increased ability in averaging away the original background noise. We present breakeven conditions that delineate when DD is useful, and further find that there is a limit in the performance of concatenated DD, specifically in how far one can concatenate the DD pulse sequences before the added noise no longer offers any further benefit in error mitigation.
\end{abstract}

\maketitle

\section{Introduction}
The adverse effects of noise pose some of the biggest challenges in realizing useful quantum technologies. The very quantum effects that give quantum technologies their edge over classical devices are also the obstacles to success: They are extremely fragile and easily destroyed by the presence of unwanted interactions with the environment noise. Much of the current research and technological push in the community are centered around exploring ways to reduce and remove the effects of noise in quantum devices~\cite{ladd2010quantum,preskill2018quantum}.

An effective noise-suppression technique is dynamical decoupling 
(DD), which requires the application of fast control pulse sequences on individual qubits to average away the effects of noise processes~\cite{
viola1998dynamical,*viola1999universal,*viola1999dynamical, 
zanardi1999symmetrizing,
khodjasteh2005fault,*khodjasteh2007performance,
viola2005random,
uhrig2007keeping,uhrig2009concatenated,
west2010near,
quiroz2011quadratic,
wang2011protection}.
Advancing well beyond its root in control techniques for NMR systems, DD has been used in many different types of experiments as a viable way to combat decoherence in quantum information processing systems~
\cite{
biercuk2009optimized,*biercuk2009experimental,
delange2010universal,medford2012scaling,zhang2014protected,farfurnik2015optimizing,wang2016experimental,merkel2021dynamical}, alongside other applications such as noise spectroscopy~\cite{szankowski2017environmental} and quantum metrology~\cite{sekatski2016dynamical}. 
Compared with  quantum error correction
(QEC)~(see, for example, Ref.~\cite{nielsen2010quantum}), a more widely studied noise-removal approach, DD is much more economical as it
requires no encoding of logical qubits using multiple physical qubits, nor real-time close-loop control through periodic syndrome measurement and recovery. All that is needed are regular single-qubit fast pulses that are usually easy to implement---they employ the same gates used for quantum computational tasks that are typically part of the capabilities of the quantum device. DD can be used by itself, or as the first layer of defense against noise within a standard QEC scheme as a hybrid noise-reduction approach~\cite{khodjasteh2009dynamically,ng2011combining,paz-silva2013optimally}.
The use of DD does not, however, come at no cost. The multiple pulses that have to be applied can be imperfect. Imperfect DD pulses can add, rather than remove, errors in the system. When the pulses are too noisy, those added errors can happen often enough to eliminate the benefit of having DD in the first place~\cite{ng2011combining}.

Like the concatenated codes in QEC, it is also possible to construct pulse schemes in a recursive manner to ``scale up'' the power of  decoupling and form what is known as concatenated DD (CDD)~\cite{khodjasteh2005fault,*khodjasteh2007performance}.  In an ideal world, one could in principle construct arbitrarily accurate DD-protected gates through concatenation~\cite{khodjasteh2010arbitrarily}. However, under realistic constraints such as control errors and finite pulse rate, both theoretical papers and experimental papers find that increasing concatenation may not always be beneficial~\cite{khodjasteh2010arbitrarily,zhang2007dynamical,*zhang2008longtime,alvarez2010performance,hodgson2010optimized,khodjasteh2011limits,piltz2013protecting,liu2013noiseresilient}. 

In this work, we study the efficacy of DD as a noise-removal technique when the DD pulses themselves are noisy by asking similar fault-tolerance questions usually asked of QEC procedures. Our investigation can be split into two lines of inquiry. First, for a given DD sequence, what is the maximum amount of noise in the DD pulses that can be tolerated, before DD stops offering any benefit? We refer to this maximum amount of tolerated noise as the breakeven point (it is also sometimes referred to as the pseudothreshold in QEC and fault-tolerance literature). Second, we ask for the accuracy threshold---again borrowing terminology of QEC---specifying the level of noise permissible in the DD pulses below which a given prescription for scaling up the DD scheme can remove more and more noise, and hence attain better and better computational accuracy in the quantum device. As we will see, we find that, in typical situations, there is no nonzero threshold. Instead, there is a maximal scale for DD, beyond which no additional benefit can be derived.

Past literatures involve some aspects of our queries. In particular, the performance of DD with imperfect pulses has been extensively studied with various models accounting for the effects of finite pulse width and systematic unitary rotation errors \cite{zhang2008longtime,west2010high,souza2011robust,souza2012robust,bernad2014effects}.
A generic description of noisy DD has also been attempted using a stochastic process  model \cite{bernad2014effects}. 
Experimental evidence suggesting the adverse impacts of noisy controls are also available in many reports~\cite{alvarez2010performance,aliahmed2013robustness,lang2019nonvanishing}.
Among these studies on non-ideal DD, however, a comprehensive cost-benefit analysis from the fault-tolerance perspective is yet available to the best of our knowledge. 
Furthermore, for a fully fault-tolerant discussion relevant to experiments today, more general noises for the pulses have to be incorporated. Experimentally, gate-control noise, which can arise every time a DD pulse is applied, is often of a very different nature than the background noise---assumed in the case of finite pulse widths---in a quantum device, and can be highly dependent on the specific gate being applied. Such noise has to be treated separately in a realistic study of the efficacy of DD.

Here, we focus our discussion on the most commonly used scheme of periodic DD (PDD) based on what is known as the universal decoupling sequence  \cite{viola1999dynamical}. The same analysis, however, extends to other DD schemes. For scaling up the DD protection, to be able to remove more noise, we make use of CDD \cite{khodjasteh2007performance}, organizing the DD pulse sequence in a recursive manner.
We discuss the breakeven point in two operational settings: computation and memory. In the computational setting, we consider using DD to reduce noise in computational gates in the course of carrying out a quantum circuit. The breakeven comparison is thus about the noise per gate, with or without DD. In the memory setting, we instead compare the noise over a fixed time interval, during which DD can be carried out or not. Most of our discussion will be in the computational setting, the one most relevant to the current interest in quantum devices, though we mention the memory setting in specific cases. 
Following past papers, we quantify the performance of DD using the error phase, namely, the strength of the effective noise Hamiltonian---with and without DD---acting on the qubits in the quantum device. As we will see, the error phase permits easy analytical treatment. We numerically check for consistency with another natural figure of merit, the system state infidelity, in a specific physical setting. 

Below, we begin in Sec.~\ref{sec:Prelim} with a few basic concepts needed for our analysis. Section \ref{sec:PDD} discusses the breakeven point for PDD, starting with the ideal-pulse situation before moving to the realistic noisy case. We examine the example of unitary errors in detail, numerically exploring the infidelity measure in addition to the analytical treatment of the error phase. Sec.~\ref{sec:CDD} extends our discussion to the case of CDD and explores the existence of an accuracy threshold for DD. We conclude and summarize our findings in Sec.~\ref{sec:Conc}.

\section{Preliminaries}\label{sec:Prelim}
We begin our discussion with the introduction of a few basic concepts necessary for understanding the rest of the paper.

\subsection{Basics of DD}
\begin{figure}[ht!]
 \includegraphics[width=1\linewidth]{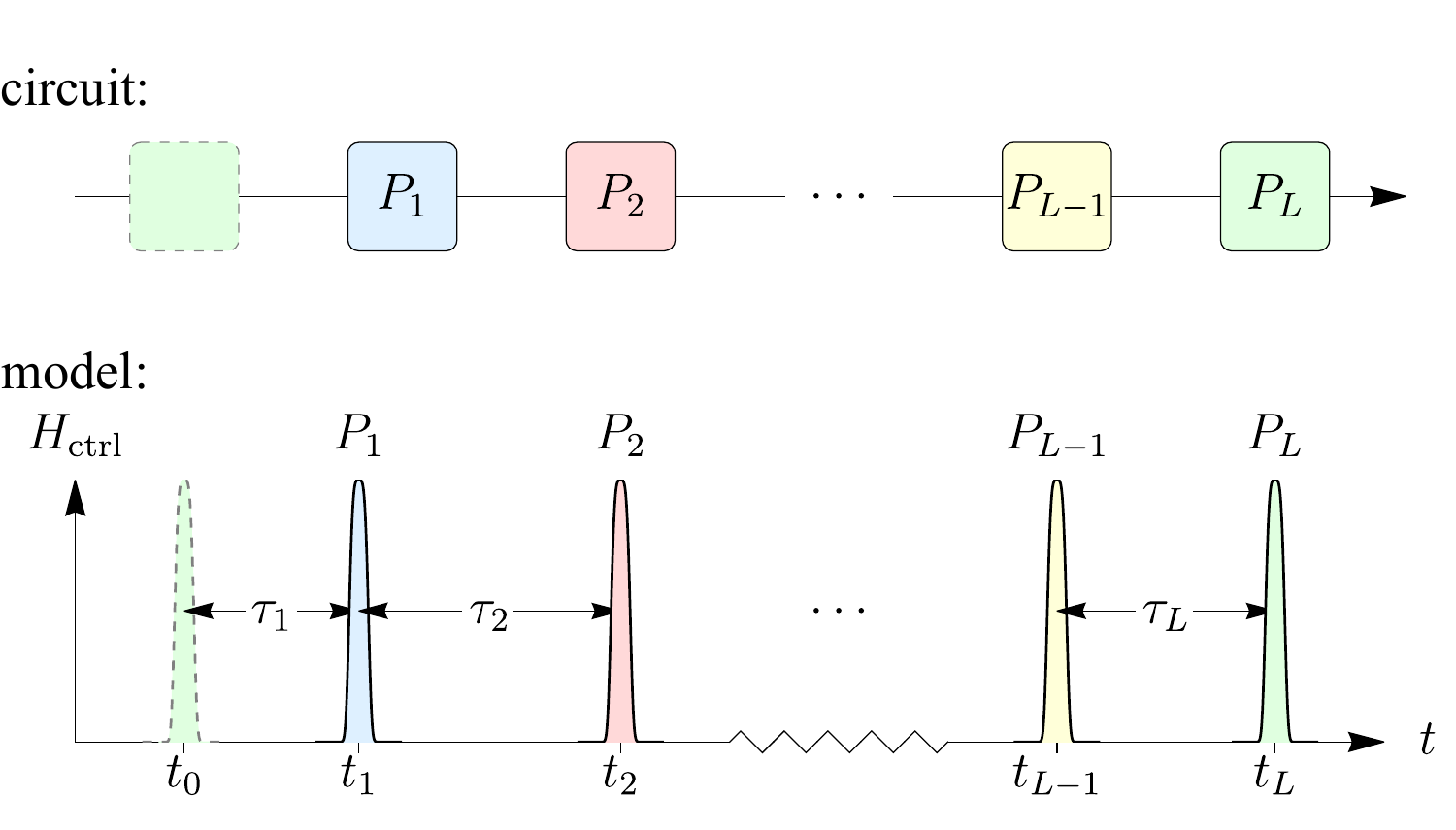}
 \caption{Illustration of one cycle of a generic DD sequence comprising $L$ pulses $P_i$, for $i=1, 2,\ldots, L$. Pulse $P_i$ is applied at time $t_i$; pulses are separated by time $\tau_i$. The dashed pulse at time $t_0$ is the final pulse of the previous DD cycle. $H_\textrm{ctrl}$ is the control Hamiltonian that implements the DD pulses.}
 \label{fig:pulses}
\end{figure}

DD involves the repeated application of a fixed sequence of short pulses (or fast gates) to individual quantum registers that average away the effect of any noise with a time scale slow compared to the sequence time. For a given DD scheme, let $L$ be the length, i.e., the number of pulses, of the sequence. We denote the $i$th pulse of the sequence as $P_i$, and write the sequence as $P_L \ldots P_2P_1$, proceeding from right to left in time. Pulse $P_i$ is applied at time $t_i=t_{i-1}+\tau_i$, for $i=1,2,\ldots, L$, with $\tau_i$ the time between pulses $P_{i-1}$ and $P_i$, and $t_0$ is the time of the last pulse of the previous sequence, coincident with the start time of the current sequence; see Fig.~\ref{fig:pulses}.

All DD schemes share some fundamental similarities. 
First, all DD pulses are chosen from a specific transformation group.
Second, all DD sequences must satisfy the constraint that in the absence of errors, $ P_L P_{L-1} \cdots P_1$ is the identity map (up to a phase factor). This constraint ensures that there is no net transformation on the quantum register at the completion of the sequence. Different strategies could differ in the following respects:
(i) the transformation group.---For a qubit register, a common choice is the Pauli group generated by the Pauli operators $X$ and $Z$. Simpler schemes such as spin echo \cite{hahn1950spin} and the CPMG sequence \cite{carr1954effects,*meiboom1958modified} use only the subgroup $\{I,Z\}$.  
(ii) The  specific sequence of pulses, which can be deterministic as well as randomized~\cite{viola2005random}.
(iii) The pulse times.---One can have regular-interval pulses, with $\tau_i\equiv \tau$ $\forall i$, as is the case in PDD \cite{viola1999dynamical} and CDD \cite{khodjasteh2005fault} schemes. One could, however, have variable-interval schemes, such as the Uhrig DD sequence and its variants \cite{uhrig2007keeping,wang2011protection,kuo2011quadratic}.
It is also possible to apply continuous control that follows a Eulerian cycle instead of using the ``bang-bang'' style control~\cite{viola2003robust}.
In this work, we focus on the regular-interval schemes of PDD and CDD, differing in sequence length and the specific pulse sequence, but which employ pulses drawn from the Pauli group.  The physical model we are considering is one where gates, including the DD pulses, are applied at some operating frequency, with time $\tau$ in between consecutive gates. Such nonzero $\tau$ can simply be the finite switch time between gates, or there may be other practical reasons for a synchronized clock cycle time.

Generally speaking, the errors can be attributed to two major sources:  
(i) the system (quantum register) interacting with its environment (local quantum bath and field fluctuations, etc.), which constitutes the background noise on the system presenting even in the absence of control; 
(ii) the control imperfections. In many past works, imperfect DD pulses were modeled as finite-width (or finite-duration) pulses during which the always-on system-bath interaction acts and leads to errors in the pulses.  
Our approach, explained in Sec.~\ref{sec:noiseModel}, allows additionally incorporating general control errors in the pulses as well. This second source of errors is perhaps more dominant and relevant in practical situations. 
In the followings, we refer to the DD sequences with perfect, instantaneous pulses as ideal DD, and to the case with imperfect pulses as noisy DD.

In the absence of any DD pulses, the system and bath evolve jointly according to the Hamiltonian $H$, assumed to be time-independent as is appropriate for standard DD analysis \footnote{Put differently, any parametric time dependence in the joint dynamics occurs because of degrees of freedom excluded from the bath; here we think of all such degrees of freedom as part of the bath.}. $H$ here can be written as $H=\HB +\HSB$, with $\HB$ as the bath-only Hamiltonian, and $\HSB$ the interaction. No system-only term appears in $H$ as we assume no nontrivial dynamics, other than that arising from $\HSB$, occur in the system during the course of a DD sequence. Computational gates, if any, can only be done in between complete DD sequences for the noise averaging to work. 
With this, we can write the evolution operator for a single complete ideal DD sequence as
\begin{equation}\label{eq:Udd}
U_\DD = P_L\upe^{-\upi\tau H}P_{L-1}\ldots P_2\upe^{-\upi\tau H}P_1\upe^{-\upi\tau H}\equiv\upe^{-\upi\Odd}.
\end{equation}
Here, we have defined $\Odd\equiv T\Heff $ as the dimensionless effective Hamiltonian appropriate for describing the evolution for the time $T\equiv L\tau$ of the DD sequence.
$\Odd$ can be written formally using the Magnus expansion,
\begin{equation}\label{eq:Oeff Magnus series}
\Odd  = \sum_{m=1}^\infty \Odd\up{m},
\end{equation}
where the $m$th term consists of products of $m$ copies of $\tau H$, and hence is of order $\Vert \tau H \Vert^m$. 
We refer the reader to past analyses of DD (see, for example, Ref.~\cite{ng2011combining}) for a detailed derivation of the Magnus expansion. Here, we provide only the basic expressions needed for our discussion below. In particular, we will need the expressions for the three lowest-order Magnus terms for piecewise-constant Hamiltonian evolution. For $U=\upe^{-\upi O_K}\upe^{-\upi O_{K-1}}\ldots\upe^{-\upi O_1}$, for $O_i$s a sequence of time-independent dimensionless Hamiltonians (e.g., $O_1=\tau H$), we have $U\equiv\upe^{-\upi\Omega}$ with $\Omega = \Omega^{(1)}+\Omega^{(2)}+\Omega^{(3)}+\ldots$, where
\begin{align}
\Omega^{(1)}&=\sum_{i=1}^K O_i,\notag\\
\Omega^{(2)}&=-\frac{\upi}{2}\sum_{\substack{i,j=1\\ i>j}}^K\, [O_i,O_j],\\
\textrm{and}\quad \Omega^{(3)}&=-\frac{1}{6}\!\!\sum_{\substack{i,j,k=1\\i\geq j\geq k}}^K\!\!\!\frac{\bigl([O_i,[O_j,O_k]]+[O_k,[O_i,O_j]]\bigr)}{\operatorname{sym}(i,j,k)}.\notag
\end{align} 
Here, $\operatorname{sym}(i,j,k)$ is the symmetry factor
(equaling to $1$ when $i,j,k$ are all different, $2$ when any two indices are equal and $6$ when all indices are equal); $[\,\cdot\,,\,\cdot\,]$ denotes the commutator.
Within the radius of absolute convergence of the Magnus series~\cite{blanes2009magnus}, 
\begin{equation}\label{eq:Magnus-abs-converge}
L\norm{\tau H }  < 1.0868\cdots \approx 1,
\end{equation}
$\Omega^{(m)}$ decreases in importance as $m$ increases. A DD scheme such that $\Omega_\DD^{(m)}$ acts trivially on the system (i.e., acts as the identity on the system) for all $m\leq n$ is said to achieve $n$th-order decoupling. For such a scheme, the system sees an effectively weakened noise, of strength $\Vert\Omega_\DD^{(n)}\Vert\sim\Vert \tau H\Vert^n$, compared with $\Vert \tau H\Vert$ without DD.

\subsection{The PDD scheme}
We specialize here to the case of interest, that of the single-qubit PDD scheme based on the universal decoupling sequence \cite{viola1999dynamical}. We refer to this case simply as PDD for brevity. The single qubit interacts with a bath, via a joint Hamiltonian (in the absence of DD) that can be written, without loss of generality, as
\begin{equation}\label{eq:H}
H\!\equiv \!I\otimes B_I+X\otimes B_X+Y\otimes B_Y+Z\otimes B_Z.
\end{equation}
Here, $I\equiv\sigma_0$ is the identity operator on the qubit, $X\equiv \sigma_1,Y\equiv\sigma_2$, and $Z\equiv \sigma_3$ are the Pauli operators on the qubit, and the $B_\alpha$s are operators on the bath. We identify $I\otimes B_I$ as the bath-only Hamiltonian $\HB$, and $X\otimes B_X+Y\otimes B_Y+Z\otimes B_Z$ as the interaction Hamiltonian $\HSB$.

The universal decoupling sequence refers to a simple 4-pulse sequence `$Z-X-Z-X-$'. The corresponding $\Omega_\DD$ of Eqs.~(\ref{eq:Udd}), now re-labeled as $\Omega_\PDD$, is expressible as the series $\Omega_\PDD=\Omega_\PDD^{(1)}+\Omega_\PDD^{(2)}+\ldots$ whose first two terms can be worked out to be
\begin{align}
\Omega_\PDD^{(1)}&=(4\tau) I \otimes B_I \label{eq:PDDMag1},\\
\Omega_\PDD\up{2}&=-(2\tau^2)\bigl\{ X\otimes 2\upi\,[B_I,B_X]\label{eq:PDDMag2}\\
&\!\!\quad\qquad +Y \otimes {\left(\upi\,[B_I,B_Y] + \{B_X,B_Z\}\right)}\bigr\},\nonumber
\end{align}
where $\{\cdot,\cdot\}$ is the anti-commutator. 

Let us compare the PDD evolution with the evolution without DD over the same time period of $4\tau$ (i.e., in the memory setting): $U=\upe^{-\upi4\tau H}\equiv \upe^{-\upi\Omega}$, with
\begin{equation}
\Omega=(4\tau)H=(4\tau){\left(I\otimes B_I+\HSB\right)}.
\end{equation}
Comparing $\Omega_\PDD^{(1)}$---usually the dominant term---with $\Omega$, we see that the $\HSB$ in $\Omega$ no longer appears in $\Omega_\PDD^{(1)}$ and $\Omega_\PDD^{(1)}$ is trivial on the system. This corresponds to the fact that PDD is able to remove the lowest-order noise and that it achieves first-order decoupling.

\subsection{Scaling up the protection with concatenation}\label{sec:PrelimCDD}
A given DD sequence yields a given decoupling order, setting a limit on the scheme's ability to reduce noise in the system. To increase the power of the DD scheme, one can employ the method of concatenation introduced in Ref.~\cite{khodjasteh2005fault}. In that work, CDD was built upon the basic PDD scheme; the same procedure of concatenation, however, can be applied to other basic DD sequences as well~\cite{uhrig2009concatenated,west2010near}. The idea is to make use of concatenation to increase the decoupling order of the resulting DD sequence, hence reducing the residual noise.

CDD can be described in a recursive manner. We begin with the bare evolution, without any DD sequence, writing the evolution operator over time interval $t$ as
\begin{equation}
U_0(t)\equiv \upe^{-\upi t H_0},
\end{equation}
where $H_0\equiv H$, with the subscript $0$ added here in preparation for concatenation to higher levels. With a DD scheme of $L$ pulses, applied at time intervals (assuming regular-interval DD) $\tau_0$, the evolution operator is
\begin{equation}
U_1(\tau_1)\equiv P_LU_0(\tau_0)\ldots P_2U_0(\tau_0)P_1U_0(\tau_0),
\end{equation}
where $\tau_1\equiv L\tau_0$. The subscript $1$ is to be understood as indicating that this is for concatenation level 1. To concatenate further, $U_k(\tau_k)$ is defined recursively, 
\begin{equation}
\begin{aligned}
U_{k}(\tau_{k})&\equiv P_L U_{k-1}(\tau_{k-1}) \ldots P_1 U_{k-1}(\tau_{k-1})\\
 &\equiv \upe^{-\upi\tau_k H_k}=\upe^{-\upi\Omega_{\CDD{k}}} \qquad (\text{for }k\ge1),
\end{aligned}
\end{equation}
with $\tau_{k}\equiv L\tau_{k-1}$. For each $U_k(\tau_k)$, we also associate an effective Hamiltonian $H_k$, and a dimensionless Hamiltonian $\Omega_{\CDD{k}}\equiv\tau_kH_k$.
Each CDD scheme is determined by specifying the maximal concatenation level $n$, and either the value of $\tau_0$ or $\tau_n$. CDD at level $n$, denoted as $\CDDn$, is then a sequence of $L_n\equiv L^n$ pulses separated by time interval $\tau_0$ and taking total time $\tau_n=L^n\tau_0$ to complete.

In the remainder of the paper, we will restrict our discussion to CDD built upon the basic PDD scheme. The authors of Ref.~\cite{khodjasteh2005fault} showed that $\CDDn$ achieves $n$th-order decoupling. This quantifies the benefit of scaling up the noise protection by concatenation. Appendix \ref{app:CDD} re-derives this conclusion with a different argument than that in the original reference.

\subsection{Quantifying the efficacy of DD}
We need concrete figures of merit to quantify the performance of DD. For most of the discussion, we will employ the error phase, which measures the strength of the effective noise Hamiltonian. In a specific example, we will also examine the infidelity measure, and compare the conclusions to those from error phase considerations.

\subsubsection{Error phase}\label{sec:PrelimEP}
To gauge the efficacy of DD, we quantify the deviation of the actual state of the quantum system, with and without DD, from the ideal, no-noise state. Following Ref.~\cite{khodjasteh2007performance}, we make use of the \emph{error phase}, which measures the strength of the system-bath interaction, the source of noise on the system. The system and bath evolve jointly for some specified time $T$ according to the evolution operator $U(0,T)$. The underlying joint Hamiltonian generating the dynamics can be time-dependent, and can include---or not---the DD pulses on the system. We write $U(0,T)\equiv \upe^{-\upi \Omega}$, for some effective dimensionless Hamiltonian $\Omega$; this can be thought of as evolution according to a time-independent effective Hamiltonian $\Omega/T$, for time $T$. $\Omega$ can be split into two pieces: $\Omega\equiv \Omega_\rB+\Omega_\rSB$, where $\Omega_\rB\equiv \frac{1}{d_\rS}I_\rS\otimes\tr_S(\Omega)$ ($d_\rS$ is the dimension of the system) acts on the bath alone, while $\Omega_\rSB\equiv \Omega-\Omega_\rB$ contains all the pieces that act nontrivially on the system. $\Omega_\rSB$ can be thought of as the effective system-bath interaction over this time $T$. We define the error phase $\Phi_\rSB$ as the norm
of $\Omega_\rSB$, $\Phi_\rB$ as the norm of $\Omega_\rB$, which as we will see, will also enter our analysis:
\begin{equation}\label{eq:ErrorPhase}
\Phi_\rB \equiv \norm{\Omega_\rB}\quad \text{and} \quad\ep\equiv \Vert\Omega_\rSB\Vert .
\end{equation}
We choose to employ the operator norm, i.e., the maximal singular value of an operator, for all norm symbols appearing in this paper.
For this choice, we have unit norm for the identity operator in any dimension, so that the bath dimension is irrelevant in the definition of the error phase, a fact that will come in useful later. 
In what follows, we will use $\phi$ for the bare Hamiltonian associated with the no-DD situation. Specifically, we write,
\begin{align}
\phiB\equiv\Vert\tau\HB\Vert\quad\textrm{and}\quad \phiSB\equiv\Vert\tau\HSB\Vert,
\end{align}
and exclusively reserve the uppercase $\Phi$ for the effective Hamiltonian after DD.
The Hamiltonian operators are assumed to be bounded throughout this work, but there is otherwise no restriction on the dimension of the bath Hilbert space.

\subsubsection{Infidelity measure}\label{sec:PrelimInFid}
We mention another natural measure of noise, namely, the infidelity between the noisy (with or without DD) and ideal no-noise system states. Since we are interested only in how the system state is affected by noise, we care only about the effective noise channels $\cN$ acting on the system, with or without the DD pulses, with the bath degrees of freedom discarded, i.e.,
\begin{align}
\cN(\,\cdot\,)&\equiv \tr_\mathrm{B}{\left\{\upe^{-\upi\Omega}(\,\cdot\,\otimes\rho_B)\upe^{\upi\Omega}\right\}}\\
\text{and}\quad\cN_\DD(\,\cdot\,)&\equiv \tr_\mathrm{B}{\left\{\upe^{-\upi\Omega_\mathsf{DD}}(\,\cdot\,\otimes\rho_B)\upe^{\upi\Omega_\mathsf{DD}}\right\}}.\nonumber
\end{align}
With these channels, we can define our infidelity measure as
\begin{align}
\infid &\equiv \max_{\cN,\psi}\infid(\cN,\psi),\label{eq:trdisInFDef}
\end{align}
with $\infid(\cN,\psi)\equiv \sqrt{\langle\psi|(\cI-\cN)(\psi)|\psi\rangle}$, noting that $\langle\psi|(\cI-\cN)(\psi)|\psi\rangle=1-\langle\psi|\cN(\psi)|\psi\rangle\in[0,1]$, so that $\infid$ and $\infid(\cN,\psi)\in[0,1]$.
We have an analogous expression for $\infid_{\DD}$ computed from $\cN_{\DD}$.
Here $\psi\equiv |\psi\rangle\langle \psi|$ is a pure system-only state. $\infid$ is the square root (as we will see, the square root gives the proper comparison with the error phase) of the deviation from 1 of the square of the fidelity between the post-noise state and the initial state. The maximization over $\cN$ (and $\cN_\DD$) refers to a maximization over all choices of the $B$ operators that enter $H=\HB+\HSB$, with fixed $\phi_\mathrm{B}$ and $\phi_\mathrm{SB}$ values. The maximization---resulting in a worst-case measure---over all pure system states provides a state-independent quantification, while the maximization over the $\cN$s reflects our typical lack of knowledge of the precise forms of the $B$ operators, even if we are given the $\phi_\mathrm{B}$ and $\phi_\mathrm{SB}$ values.


One can obtain analytical bounds between the error phase and infidelity measures. Appendix \ref{app:epInF} derives such a bound by writing the output state $\cN(\rho_\mathrm{S})$ as a power series in $\Omega$. We find the relation
\begin{equation}
0\leq \infid(\cN,\psi)\lesssim \sqrt2\phi_\mathrm{SB}.
\end{equation}
Since we allow any general channel $\cN$ here, analogous bounds also hold for $\infid_\DD$ and $\Phi_\mathrm{SB}$.

\section{Limits of decoupling: the breakeven conditions}\label{sec:PDD}
Any noise mitigation strategy is effective only if its costs, i.e., the increased complexity of carrying out the computational task with noise mitigation, are lower than its benefits, i.e., the increased ability to reduce the adverse effects of noise on the computation. The costs enter not just because any noise mitigation approach requires the use of more resources, e.g., more gates, more qubits, etc., but also that those added resources are themselves imperfect in practice, such that the unmitigated noise of the resulting larger system is unavoidably larger than if no noise mitigation strategy was adopted. There is then a net benefit only if the added noise is low enough to not overwhelm the added noise-removal capabilities. Such is the content of any fault-tolerance analysis (see, for example, Ref.~\cite{nielsen2010quantum}), usually applied to quantum computing tasks protected by quantum error correction. Here, we apply the same logic to DD, and ask when the added benefit of averaging away part of the noise outweighs the added cost of having to do additional pulses which are themselves noisy.

For a proper cost-benefit analysis, we distinguish between two operational scenarios: to better preserve states in a quantum memory (no computational gates), or to reduce noise in the course of carrying out a quantum computation. In the memory setting, the goal is to preserve an arbitrary quantum state for some storage time $T$. During this period, DD pulses are applied, and we can ask if the resulting noise is lower compared with the no-DD case over the same time period.
In other words, for DD to work well, we require
\begin{equation}
\textrm{Memory setting:}\quad 
\epsilon_\DD(T)<\epsilon(T),
\end{equation}
where $\epsilon(t)$ is some chosen figure of merit quantifying the noise associated with the time evolution over period t, with or without DD as specified by the subscript. In particular, for the quantum memory protected with an $L$-pulse DD scheme with constant interval $\tau$, we have $T=L\tau$.
In the computational setting, instead of the same total evolution time with and without DD, we need to pay attention to the gate time, assumed to be common to both computational and DD gates. In the absence of DD, we assume computational gates are applied at the same rate as DD pulses, i.e., every time step $\tau$. With DD, however, the protected computational gates are further separated in time by a factor $L$, the number of pulses in the DD cycle.
The condition for DD to be effective in this case is then
\begin{equation}\label{eq:condComp}
\textrm{Computational setting:}\quad
\epsilon_\DD(L\tau)<\epsilon(\tau).
\end{equation}
In our work, we are more interested in the computational setting, the more stringent one among the two, though it is straightforward to adapt our analyses to the memory case, as we do in some of the situations below. As we will see, Condition \eqref{eq:condComp} will yield a requirement on the noise parameters characterizing the noise in the system and the DD pulses. 
We refer to this requirement as the breakeven condition for DD, borrowing terminology from quantum error correction and fault-tolerant quantum computing.

\subsection{Ideal case}\label{sec:PDDBreakEven}
Let us first discuss the breakeven condition for ideal PDD. We expect to recover the usual statement of when DD works at all, namely when the noise changes slowly compared with the time taken for a complete DD sequence. We choose the error phase as our figure of merit: $\epsilon\equiv \Phi$. For PDD to be useful, we require, under the computational setting,
\begin{equation}\label{eq:condition-ideal}
\Phi_\rSB\leq \phiSB,
\end{equation}
where $\Phi_\rSB$ is the error phase with PDD, while $\phiSB$, as defined earlier, is $\Vert\tau\HSB\Vert$, the error phase without DD, for a single time step $\tau$.
From our earlier discussion of PDD, we have $\ep=\Vert \Omega_\PDD^{(2)}+\Omega_\PDD^{(3)}+\ldots\Vert$. The full Magnus series is difficult to write down, but we can employ the dominant nontrivial (on the system) term, $\Omega_\PDD^{(2)}$, for the approximate condition:
\begin{equation}\label{eq:cond1}
\ep\simeq\Vert \Omega_\PDD^{(2)}\Vert\leq \phiSB.
\end{equation}
Using the expression for $\Omega_\PDD^{(2)}$ from Eq.~\eqref{eq:PDDMag2}, we have
\begin{align}
\Vert\Omega_\PDD^{(2)}\Vert&\leq 2\tau^2{\bigl(2\Vert[B_I,B_x]\Vert+\Vert[B_I,B_Y]\Vert +\Vert\{B_X,B_Z\}\Vert\bigr)}\nonumber\\
&\leq 4\phiB\tau\bigl(2\Vert B_X\Vert +\Vert B_Y\Vert\bigr) +4\tau^2\Vert B_X\Vert\Vert B_Z\Vert\nonumber\\
&\leq 12\phiB\phiSB+4\phiSB^2\label{eq:Opdd2}\,,
\end{align}
noting that $\phiB\equiv \Vert\tau H_\mathrm{B}\Vert =\Vert \tau B_I\Vert$, and that $\phiSB\equiv \Vert\tau H_\mathrm{SB}\Vert\geq \Vert\tau B_i\Vert$ for $i=X,Y,Z$ (see App.~\ref{app:NormIneq}).
For the breakeven condition \eqref{eq:condition-ideal}, it then suffices to require
\begin{equation}\label{eq:PDDCond}
12\phiB+4\phiSB\leq1.
\end{equation}

Condition~\!\eqref{eq:PDDCond} amounts to a requirement that both $\phiSB$ and $\phiB$ be small for PDD to work well. We note that this condition is also sufficient for the convergence criterion, Eq.~\!\eqref{eq:Magnus-abs-converge}, as $4\norm{\Omega}\le 4\phi_\rB + 4\phi_\rSB \le 1$, which in turn justifies the approximation of Eq.~\eqref{eq:cond1}.
The coefficients for $\phi_\rB$ and $\phi_\rSB$ contain a factor of 4 from the length of the PDD sequence. In general, a longer sequence will have a larger pre-factor and hence put a more stringent requirement on the noise to be small.
 That $\phiSB$ has to be small comes as no surprise: $\phiSB$ quantifies the strength of the noise on the system, and DD is expected to work well as long as the noise is weak enough so that the remnant noise is small. The requirement that $\phiB$ be small is perhaps more surprising. After all, it bounds the bath-only term which does not directly lead to noise on the system. Nevertheless, $H_\mathrm{B}$ determines the evolution rate of the bath (which can be made explicit in the Heisenberg picture of $H_\rSB$ defined by $H_\rB$), while $\tau$ is the inverse rate for the control pulse.
Hence $\phiB$ quantifies how rapidly the bath evolves relative to the pulse, or how fast the noise is compared with the control.
 That $\phiB$ should also be small should be understood as the additional requirement that the characteristic frequency of the bath should be low compared with the control, for good averaging over the entire sequence. These two requirements are consistent with the understanding of DD from the filter-function formalism~\cite{szankowski2017environmental}.

\subsection{Noisy DD gates}
Next, we accommodate the possibility of noise in the application of the DD pulses. A realistic---imperfect or noisy---DD pulse takes finite time to complete, during which the always-on background $\HSB$ interaction acts, leading to noise in the applied pulse. Control errors in the course of the gate application contribute to additional imperfections.

\subsubsection{Noise model}\label{sec:noiseModel}
We describe a noisy DD pulse in the following manner. We denote the actual $i$th DD pulse by $\widetilde\cP_i$, now regarded as a \emph{map} on states, not just a unitary operator. $\widetilde\cP_i$ can be obtained in an experiment through the use of process tomography methods, giving generally a completely positive (CP) and trace-preserving (TP) map~\cite{breuer2007theory}. $\widetilde\cP_i$ can in reality be time-dependent in that the noisy pulse may be different in each DD cycle, but for simplicity, we assume no such time dependence. Each pulse is assumed to take time $\tau_P$ (the pulse width) to complete, during which a gate Hamiltonian $H_i$ acts. In the ideal situation, $P_i=\upe^{-\upi\tau_PH_i}$; in reality, during the gate action, the background Hamiltonian $H=\HB+\HSB$ acts as well, and there can be additional gate control noise. We hence write the noisy pulse as
\begin{equation}\label{eq:Pi}
\widetilde\cP_i=\cU_i\circ \cE_i.
\end{equation}
Here, $\cU_i(\cdot)\equiv U_i(\cdot)U_i^\dagger$ with $U_i\equiv \upe^{\upi\tau_P(H_i+H)}$, is the unitary map that accounts for the finite width of the DD pulse, with noise coming from the background Hamiltonian $H$. $\cE_i\equiv \cU_i^{\dagger}\circ \widetilde\cP_i$ is a CPTP ``error" map that captures the gate control noise. $\cE_i$ acts only on the system in the typical case. Splitting $\widetilde\cP_i$ into the two maps as in Eq.~\eqref{eq:Pi} reflects the physical nature of the two noise sources and what we typically know about them: The noise arising from the background Hamiltonian scales as the pulse width, while the control noise (e.g., finite turn-on/off times, misalignment issues, etc.) often does not. While one can view $\cE_i$ (as we do below, mathematically) as one generated by a noise Hamiltonian, one often does not know that Hamiltonian source of noise but obtains $\cE_i$ from tomography alone, as a map for the time step associated with the pulse.

Now, if $\cE_i$ is a unitary map, then $\cE_i(\cdot)=V_i(\cdot)V_i^\dagger$ with $V_i$ a unitary operator on the system. We can write $V_i\equiv \upe^{-\upi\Gamma_i}$, where $\Gamma_i$ is a Hermitian operator to be viewed as the dimensionless effective Hamiltonian for the control noise in the $i$th pulse. Even for a non-unitary CPTP map $\cE_i$ (the typical case), we can ``dilate" the system to include an ancillary system such that $\cE_i$ arises from a unitary map on the system-ancilla composite, with $\cE_i$ obtained after tracing over the ancilla.
According to the Stinespring dilation theorem of a quantum channel, it suffice to incorporate at most a $d^2$-dimensional ancilla space for the $d$-dimensional system~\cite{Watrous2018Theory}. For the analytical treatment below using the error phase, we will always assume this dilation, and treat all noise as unitary $\cE_i$ with a dimensionless effective Hamiltonian $\Gamma_i$. The ancillary system needed for the dilation is treated as part of the bath. We thus write the noisy pulse finally as $\widetilde\cP_i(\cdot)=\widetilde P_i(\cdot)\widetilde P_i^\dagger$, with $\widetilde P_i$ a unitary operator on the system and bath defined as
\begin{equation}
\widetilde P_i\equiv\upe^{-\upi\tau_P(H_i+H)}\upe^{-\upi\Gamma_i}.
\end{equation}
Note that every $\Gamma_i$ can be taken to have no pure-bath term, i.e., it does not have a term proportional to $I$ on the system. Such a term gives rise only to an overall phase on the system and cannot result in observable imperfections in the pulses.

We assume that $\Vert\Gamma_i\Vert\leq \eta$ for all $i$, where $\eta$ is a dimensionless parameter that carries the meaning of a control-noise strength. In experimentally relevant scenarios, we expect $\eta$ to be small. Also, $\tau_P$ is usually chosen to be small compared to $\tau$, for fast pulses that minimize the effect of $H$ while the pulse Hamiltonian acts. We define $\delta\equiv \tau_P/\tau\ll1$.
There is no particular relation between the small parameters $\eta$ and $\delta$, and we will consider the leading-order contributions from both terms.

\subsubsection{Noisy PDD}
With noisy and finite-width pulses, the evolution of the system and bath under DD can be described using the time evolution operator,
\begin{align}
\widetilde U_\DD&\equiv \widetilde P_K\upe^{-\upi(\tau-\tau_P) H}\widetilde P_{K-1}\ldots\\
&\qquad\ldots \widetilde P_2\upe^{-\upi(\tau-\tau_P)H}\widetilde P_1\upe^{-\upi(\tau-\tau_P)H}\equiv \upe^{-\upi\widetilde\Omega_\DD}.\nonumber
\end{align}
This reduces to the ideal case of Sec.~\ref{sec:PDDBreakEven} when $\Gamma_i$ and $\tau_P$ both vanish.

For PDD, the $P_i$ sequence comprises only $X$ and $Z$ pulses. We assume that the $X$ pulses suffer the same noise each time they are applied, as do the $Z$ pulses. We write the noisy $X$ and $Z$ pulses as
\begin{align}
\widetilde Z&=\upe^{-\upi\tau_P(H_Z+H)}\upe^{-\upi\Gamma_Z}\nonumber\\
\textrm{and}\quad\widetilde X&=\upe^{-\upi\tau_P(H_X+H)}\upe^{-\upi\Gamma_X},
\end{align}
with $H_X\equiv \frac{\pi}{2\tau_P}X$ and $H_Z\equiv \frac{\pi}{2\tau_P}Z$. We can re-write the time evolution as
\begin{align}
\widetilde U_\PDD&=\widetilde Z\upe^{-\upi(\tau-\tau_P) H}\widetilde X\upe^{-\upi(\tau-\tau_P)H}\nonumber\\
&\qquad\cdot \widetilde Z\upe^{-\upi(\tau-\tau_P)H}\widetilde X\upe^{-\upi(\tau-\tau_P)H}\\
&\equiv \upe^{-\upi K_3}\upe^{-\upi\Omega_3}\upe^{-\upi K_2}\upe^{-\upi\Omega_2}\upe^{-\upi K_1}\upe^{-\upi\Omega_1}\upe^{-\upi K_0}\upe^{-\upi\Omega_0},\nonumber
\end{align}
where $\Omega_\alpha\equiv \tau\sigma_\alpha H\sigma_\alpha$ for $\alpha=0,1,2,3$, and the $K_\alpha$s are defined as $\upe^{-\upi K_0}\equiv X\widetilde X \upe^{\upi\tau_P H}$, $\upe^{-\upi K_1}\equiv Y\widetilde Z\upe^{\upi\tau_P H}X$, $\upe^{-\upi K_2}\equiv Z\widetilde X \upe^{\upi\tau_P H} Y$, and $\upe^{-\upi K_3}\equiv \widetilde Z \upe^{\upi\tau_P H}Z$. Following the calculation detailed in App.~\ref{app:NoisyPDD}, we find, to the lowest order in $\eta$ and $\phi_\textrm{SB}\delta\,(=\Vert \tau_P\HSB\Vert)$, 
\begin{equation}\label{eq:NoisyPDD}
 \begin{aligned}
K_0 &\simeq\, \Gamma_X  + \tau_P
\begin{aligned}[t]
 \bigl[ & -Y \otimes (B_Y - (2/\pi)B_Z)\\
 &\quad - Z \otimes (B_Z + (2/\pi)B_Y)\bigr],
\end{aligned}\\
K_1 &\simeq\, X\Gamma_Z X +  \tau_P 
\begin{aligned}[t]
 \bigl[ & -X \otimes (B_X - (2/\pi)B_Y) \\
 &\quad + Y \otimes (B_Y + (2/\pi)B_X)\bigr],
\end{aligned}\\
K_2&\simeq\, Y\Gamma_X Y + \tau_P 
\begin{aligned}[t]
 \bigl[ & -Y \otimes (B_Y - (2/\pi)B_Z) \\
&\quad + Z\otimes(B_Z + (2/\pi)B_Y)\bigr],
\end{aligned}\\
K_3&\simeq\,  Z\Gamma_Z Z + \tau_P 
\begin{aligned}[t]
 \bigl[ & X \otimes (B_X - (2/\pi)B_Y) \\
 &\quad + Y \otimes (B_Y + (2/\pi)B_X)\bigr].
\end{aligned}
\end{aligned}
\end{equation}
With this, we can calculate the first-order Magnus term for noisy PDD as
\begin{align}\label{eq:noisyOpdd1}
\widetilde{\Omega}_\mathsf{PDD}^{(1)}&=\sum_{\alpha=0}^3(\Omega_\alpha+K_\alpha)\\
&\simeq(4\tau)I\otimes B_I+(4/\pi)\tau_PY\otimes{\left(B_X+B_Z\right)}\nonumber\\
&\quad+\Gamma_X+X\Gamma_ZX+Y\Gamma_XY+Z\Gamma_ZZ\nonumber.
\end{align}
Unsurprisingly, we no longer have exact first-order decoupling, with deviations of order $\phi_\mathrm{SB}\delta$ and $\eta$. 

Observe that, if $\Gamma_X=\Gamma_Z$, the $\Gamma_i$ terms in Eq.~\eqref{eq:noisyOpdd1} cancel, leaving only the finite-width term that goes as $\tau_P$ as the sole deviation from exact first-order decoupling. Thus, \emph{gate-independent} control noise affects PDD only at order $\eta^2$ and higher. We can understand this as the PDD sequence averaging away also the constant---over the PDD sequence---noise that comes from the imperfect pulses, just as it does the time-independent $\HSB$. Put differently, the deviation from first-order decoupling from control noise enters only as the \emph{difference} between the $X$ and $Z$ control noise, so that we can re-write $\widetilde\Omega_\mathsf{PDD}$ as
\begin{align}\label{eq:noisyOpdd2}
\widetilde{\Omega}_\mathsf{PDD}^{(1)}&\simeq(4\tau)I\otimes B_I+(4/\pi)\tau_PY\otimes{\left(B_X+B_Z\right)}\\
&\quad+X(\Gamma_Z-\Gamma_X)X+Z(\Gamma_Z-\Gamma_X)Z\nonumber.
\end{align}
Ignoring higher-order Magnus terms, we can extract a sufficient breakeven condition,
\begin{equation}\label{eq:BE2}
(8/\pi)\phi_\mathrm{SB}\delta +4\eta\leq \phi_\mathrm{SB}.
\end{equation}
Here, we have upper-bounded $\Vert\Gamma_Z-\Gamma_X\Vert$ by $2\eta$, the most general bound when there is no particular relation between $\Gamma_Z$ and $\Gamma_X$, as would be the case for platforms where the two gates are applied by different methods. If, on the other hand, the control noise is only weakly gate-dependent, namely, that $\Vert\Gamma_Z-\Gamma_X\Vert\leq \eta'\ll\eta$, one can replace the $4\eta$ in the breakeven condition Eq.~\eqref{eq:BE2} by $2\eta'$. In this case, $\eta'$ may also be comparable to the second-order Magnus terms (e.g., compared with $\phi_\mathrm{SB}^2$), and higher order corrections should be considered and included in the breakeven condition. We explore a concrete example of this in Sec.~\ref{sec:UError}.

In the absence of gate control noise, i.e., $\Gamma_X=0=\Gamma_Z$, or, equivalently, $\eta=0$, we return to the case of finite-width pulses already discussed in past literature. In this scenario, the breakeven condition becomes simply $\delta\leq\frac{\pi}{8}$, or, in terms of $\tau_P$ itself,
\begin{equation}\label{eq:BE2eta0}
\tau_P\leq \frac{\pi}{8}\tau\simeq 0.4 \tau.
\end{equation}
In the opposite limit, if there is control noise, but pulses are instantaneous, i.e., $\tau_P=0$, we have the breakeven condition,
\begin{equation}\label{eq:BE2tauP0}
\eta\leq \frac{1}{4}\phi_\mathrm{SB}.
\end{equation}

\subsection{Example: Unitary error and cross-measure consistency}\label{sec:UError}

\begin{figure*}[htp]
\includegraphics[width=\textwidth]{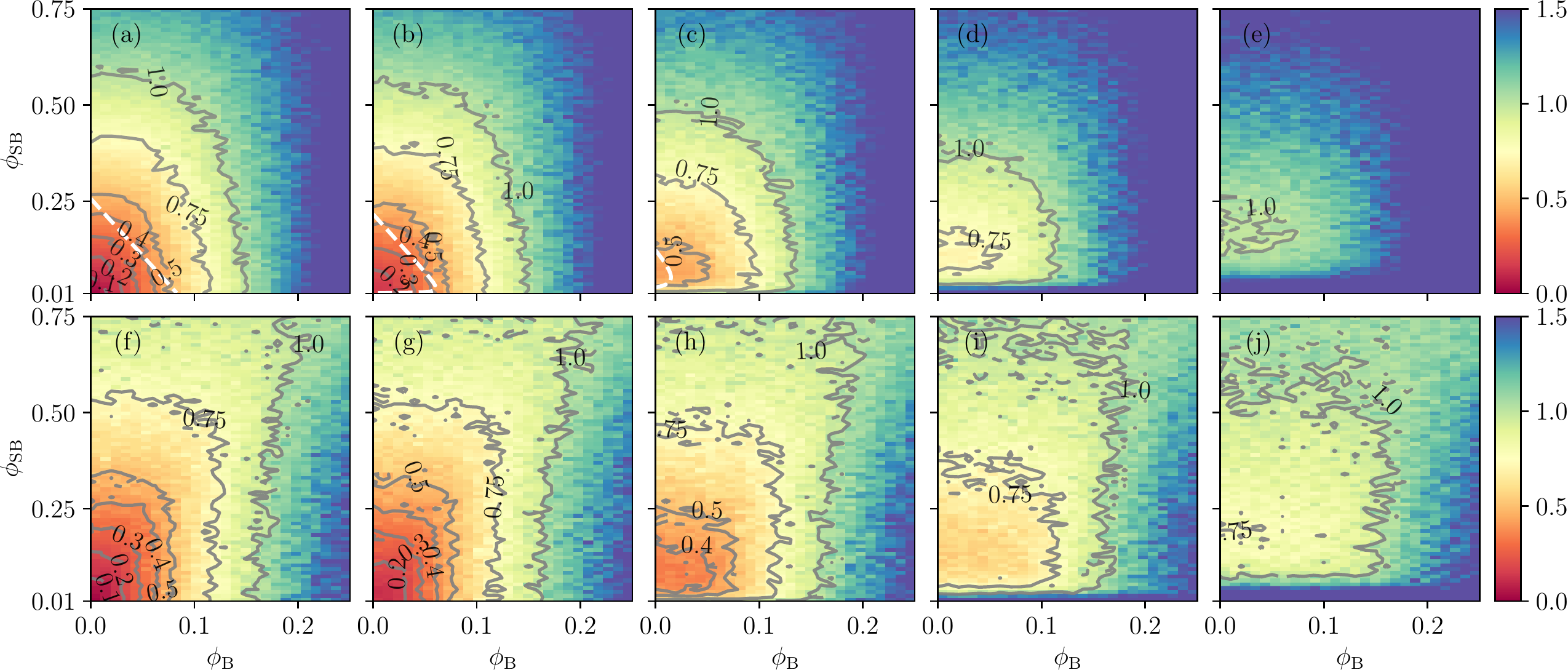}
\caption{Ratio $\epsilon_{\DD}/\epsilon(1)$ for PDD, with unitary system-only gate control noise [Eq.~\eqref{eq:uNoise}] and instantaneous pulses, for $\epsilon$ taken to be the error phase (top row), and infidelity (bottom row). The strengths of gate control noise are, from left to right, $\eta = 0,0.02, 0.06,0.1$, and $0.14$. The ratio $=1$ contours give the breakeven conditions. The white-dashed lines in (a)-(c) mark the analytical breakeven condition of Eq.~\eqref{eq:unitaryBE}.}  
    \label{fig:noisy-pulse-comparison}
\end{figure*}

As a concrete example of our discussion of breakeven conditions for PDD, let us examine the situation of a unitary---on the system only---gate control noise, arising, for example, from a systematic calibration error in the pulse control. At the same time, we investigate the infidelity measure for $\epsilon$ and $\epsilon_\DD$, as an alternative to the error phase, and compare the conclusions about the breakeven conditions. 

For simplicity, we assume instantaneous pulses $\tau_P=0$ (since finite-width errors are already explored in existing literature) and that the noise is gate independent. Specifically, we write
\begin{equation}\label{eq:uNoise}
\Gamma_X=\Gamma_Z=\eta\,\mathbf{n}\cdot\bm{\sigma},
\end{equation}
with $\bm\sigma\equiv (\sigma_X,\sigma_Y,\sigma_Z)=(X,Y,Z)$, $\mathbf{n}\equiv (n_X,n_Y,n_Z)$ is a real unit vector, and $\eta$ is a nonnegative constant quantifying the strength of the unitary error. Here, the $\Gamma$s act only on the system and not on the bath. Since $\tau_P=0$ and the noise is gate independent, the first-order Magnus contribution to the effective with-DD interaction vanishes: $\widetilde{\Omega}_\mathsf{PDD}^{(1)}$ of Eq.~\eqref{eq:noisyOpdd2} reduces to a bath-only term. The error phase is then second order in small quantities, and we need the second-order Magnus term to deduce the effective interaction Hamiltonian. Direct calculation yields,
\begin{multline}\label{eq:noisyOpdd3}
\widetilde{\Omega}_\mathsf{PDD}^{(2)} = -4X \otimes \upi\tau^2  [B_I,B_X]
- 2 Y  \otimes \bigl(\,\upi\tau^2[B_I,B_Y]  \\ 
+\{\tau B_X+\eta n_XI,\tau B_Z  +\eta n_ZI\}\bigr).
\end{multline}
We can then bound the error phase in a straightforward manner, with triangle inequalities, as
\begin{align}
\label{eq:noisyPDDbound}\Phi_\rSB\simeq\Vert\widetilde{\Omega}_\mathsf{PDD}^{(2)}\Vert&\leq 2\eta^2+8\eta\phi_\rSB+12\phi_\rB\phi_\rSB+4\phi_\rSB^2\\
&=2(\eta+2\phi_\rSB)^2+4\phi_\rSB(3\phi_\rB-\phi_\rSB),\nonumber
\end{align}
where we have used the fact that $|n_Xn_Z|\leq \frac{1}{2}$ since $\mathbf{n}$ is a unit vector.
A sufficient breakeven condition, neglecting higher-order corrections, can be obtained by restricting the above upper bound on the error phase $\Phi_\rSB$ to be no larger than $\phi_\rSB$. This gives the regions bounded by the white-dashed lines in the lower-left corners of Figs.~\ref{fig:noisy-pulse-comparison}(a)-(c); for (d) and (e) in that figure, the breakeven condition cannot be satisfied for the chosen parameters.

One can regard $\Phi_\rSB\leq \phi_\rSB$ as a threshold condition for the gate control error $\eta$, for given $\phi_\rSB$ and $\phi_\rB$. Using the  approximation in Eq.~\eqref{eq:noisyPDDbound} for the error phase, this translates into the condition
\begin{equation}\label{eq:unitaryBE}
\eta\leq \sqrt{\tfrac{1}{2}\phi_\rSB}{\left[1-4(3\phi_\rB-\phi_\rSB)\right]}^{-1/2}-2\phi_\rSB
\end{equation}
for the breakeven point.
For $\phi_\rB$ and $\phi_\rSB\ll 1$, as is usually the case in practice, this becomes $\eta\lesssim \sqrt{\phi_\rSB/2}$. The square-root relation between $\eta$ and $\phi_\rSB$ comes directly from the fact that the gate-independent noise matters only at the second order when there is DD, compared with $\phi_\rSB$ that enters in the first order without DD.

To gauge how tight our analytical bounds are, which are general and applicable for all $H$ and unitary $\Gamma$, we numerically simulate the action of PDD for specific instances to find the breakeven conditions. At the same time, numerical analysis permits the investigation of figures of merit to quantify the performance of PDD, beyond the analytically accessible error phase that we have used thus far. Our conclusions about the breakeven conditions are useful only if they are reasonably consistent across different measures. Specifically, we examine the infidelity measure, already introduced in Sec.~\ref{sec:PrelimInFid}, in addition to the error phase figure of merit.

We numerically simulate the action of PDD on a single system qubit, for the gate-independent unitary noise of Eq.~\eqref{eq:uNoise}. The bath is also taken to be a single qubit. The system-bath Hamiltonian takes the form of Eq.~\eqref{eq:H}, with the bath operators chosen randomly, with specified values of the norms, $\phi_\mathrm{B}\equiv \Vert\HB\Vert= \Vert I \otimes B_I\Vert$ and $\phi_\mathrm{SB}\equiv\Vert\HSB\Vert=\Vert\sum_i\sigma_i\otimes B_i\Vert$. $B_I$ is chosen such that $\tr(B_I)$ vanishes, corresponding to fixing a choice for the zero-energy level. Specifically, we write $B_I=\phi_\mathrm{B}\mathbf{v}\cdot\bm{\sigma}$ where $\mathbf{v}$ is a real 3D unit vector chosen uniform-randomly over the surface of a 3-sphere. In addition, we write $B_i = R_i + R_i^\dagger$, where $R_i$ is a complex random matrix with normally distributed entries. The resulting $H_\rSB$ is normalized and then multiplied by a chosen $\phi_\rSB$ to  give the desired magnitude.

The error phase approach requires no specification of the initial bath state, as it quantifies the size of the full system-bath interaction Hamiltonian. The infidelity measure that we compute here, however, targets only the resulting system-only state, and the effective system-only noise channels depend on the choice of initial bath state $\rho_\mathrm{B}$. In those cases, we consider the infinite temperature state, i.e., the maximally mixed state, as a symmetric choice of bath state. The maximization contained in the definition of $\infid$ is carried out numerically with 10,000 samples over $\cN$ and $\psi$.

In Fig.~\ref{fig:noisy-pulse-comparison}, we plot the numerical values of the ratio $\epsilon_\mathrm{PDD}/\epsilon(1)$, for the two different choices of the figure of merit $\epsilon$, over the parameter space $(\phi_\mathrm{B},\phi_\mathrm{SB})$, and for different $\eta$ values. Figs.~\ref{fig:noisy-pulse-comparison}(a-e) (top row), plotting the error phase results, show a rapid shrinking of the breakeven region, where DD remains effective, as $\eta$ increases from $0$ to $0.14$. The numerical breakeven regions are much larger than those predicted by the analytical condition of Eq.~\eqref{eq:unitaryBE}. The infidelity measure is plotted in Figs. \ref{fig:noisy-pulse-comparison}(f-j) (bottom row). With $\infid$, the breakeven regions are larger, indicating a difference in detailed conclusion from the error-phase measure. The contour shapes nevertheless follow a similar pattern as that of the error phase measure, suggesting a qualitative agreement between the two measures.

\section{Scaling up protection: limits of concatenation}\label{sec:CDD}
\begin{figure}[htbp]
 \centering
 \includegraphics[width=0.9\linewidth]{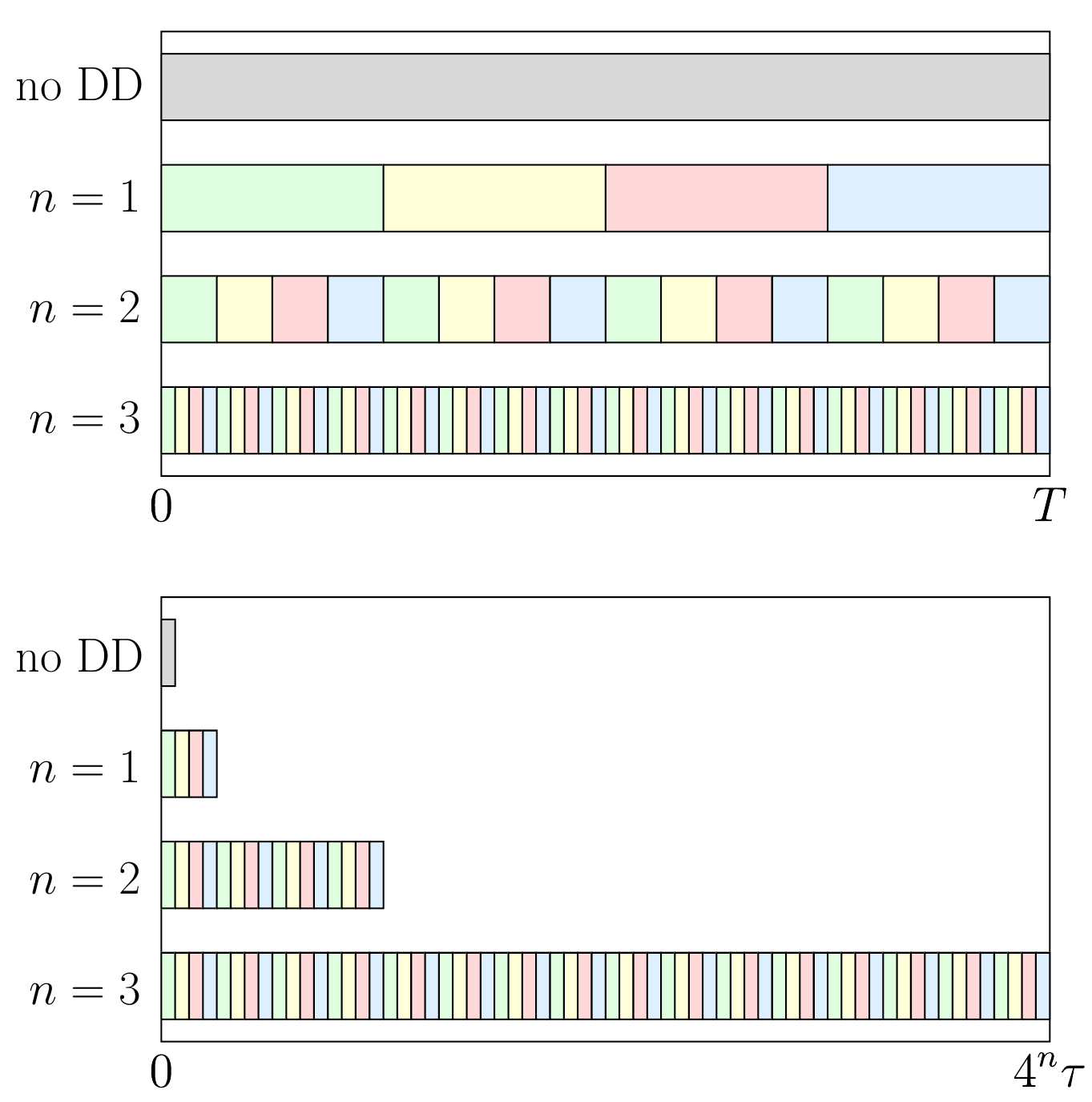}
 \caption{An illustration of the CDD sequences for different concatenation levels. We compare the memory setting (top panel) with the computational setting (bottom panel).}
 \label{fig:memvscom}
\end{figure}

Next, we turn to the situation of CDD, where we attempt to scale up 
the noise protection by concatenating a basic decoupling sequence to longer and more sophisticated sequences. The scaling strategy differs between the quantum memory scenario and the quantum computation scenario, as illustrated in Fig.~\ref{fig:memvscom}. 
In the memory setting, the natural strategy is to slice the fixed evolution time $T$ into finer and finer intervals to accommodate more and more DD pulses, until some minimal gate time is reached.
In the computational setting, with the working gate time $\tau$ fixed, 
increasing the concatenation level will inevitably bring about a longer time between computational gates. Past analytical and numerical studies gave rather optimistic assessments of the scaling performance using the memory setting \cite{khodjasteh2005fault,khodjasteh2007performance,khodjasteh2010arbitrarily,alvarez2010performance}, while 
experimental and numerical studies in the computational setting produced mixed conclusions~\cite{zhang2007dynamical,*zhang2008longtime,west2010high,piltz2013protecting}.
Below, we put our emphasis on the computational setting, the more relevant scenario for the current pursuit of high-fidelity gates for quantum computational tasks.  We also adapt our analyses to the memory case for comparison.

One can certainly discuss the breakeven point for a particular $\CDDn$ sequence, treating it simply as a ``flattened" sequence of pulses, and asking when the error phase after $\CDDn$ is smaller than $\phiSB$ as was done for PDD in the previous section. 
However, a more interesting question, given the CDD scheme of scaling up the protection by increasing the concatenation level, is the gain in noise-removing power for every increase in CDD level.
A key concept in fault-tolerant quantum computing is the accuracy threshold, the noise strength below which scaling up the QEC code always gives improved protection against noise, and hence more accurate quantum computation (see, for example, Ref.~\cite{nielsen2010quantum}). Here, we ask the analogous question of CDD: Is there a condition on the noise parameter of the problem such that increasing the CDD level always leads to improved noise removal capabilities? Concretely, an accuracy threshold for CDD exists if the physical noise strength---denoted symbolically here by $\eta$, though there can be many noise parameters that characterize the noise strength---satisfies a condition $\eta<\eta_\mathrm{thres}$, for some $\eta_\mathrm{thres}>0$, such that 
\begin{equation}\label{eq:thresCond}
\Phi_{\mathrm{SB},n+1} < \Phi_{\mathrm{SB},n},
\end{equation}
for every $n=1,2,\ldots$. Here, $\Phi_{\mathrm{SB},n}$ denotes the error phase for $\CDDn$. The existence of an accuracy threshold means that every increase in CDD level is accompanied by a decrease in the resulting error phase, as long as the physical noise strength is below the threshold level. Note that Eq.~\eqref{eq:thresCond} automatically implies the breakeven condition by recognizing that $\Phi_{\rSB,0}=\phi_\rSB$.

Unfortunately, as we will show below, for CDD built from concatenating the PDD scheme, there is no such accuracy threshold in the computational setting, i.e., there is no nonzero $\eta_\mathrm{thres}$ for which Eq.~\eqref{eq:thresCond} holds for every $n$. Instead, we will see that the error phase initially decreases as $n$ increases but eventually, this decrease turns around, and there is a maximal concatenation level beyond which the error phase actually grows with $n$. This maximal concatenation level hence quantifies the limits to the power of CDD, for given physical noise parameters.
Below, we show this general behavior for both the ideal CDD case where pulses are perfect, and the noisy CDD case where imperfections in the pulses are allowed. 

As mentioned earlier, both $\phiB$ and $\phiSB$ have to be small for DD to offer benefits.
Depending on the relative sizes of these two terms, there are two very different regimes for understanding the CDD performance: $\phiB/\phiSB\ll 1$ and $\phiB/\phiSB\gtrsim 1$.
Reference \cite{khodjasteh2007performance}, the original work that introduced CDD, focused on the limit of $\phiB/\phiSB\gtrsim 1$ (specifically, they assume $\phiB>\phiSB$ while both are small), a limit they argue to be relevant when the system is coupled only to a small portion of the bath while the whole bath, with many degrees of freedom, has a nonzero self-interaction. The other limit of $\phiB/\phiSB\ll 1$ is also potentially of physical interest, and represents an arguably better situation for DD, where the noise evolution is of secondary importance to the actual interaction between the system and bath. Here, we re-examine the situation of $\phiB/\phiSB\gtrsim 1$ as done in \cite{khodjasteh2007performance} but now with gate-control noise. We also consider the opposite limit of $\phiB/\phiSB\ll 1$ for the ideal case; we will see that the qualitative behavior is, in fact, not that different. In that case, to obtain the leading-order behavior, we can simply assume $B_I=0$ in the physical Hamiltonian. The practical situation will likely be somewhere in between the two limits.

\subsection{Ideal case}
\subsubsection{\texorpdfstring{$\phiB/\phiSB\gtrsim 1$}{\straightphi B/\straightphi SB >1}}
We first work in the limit of $\phiB/\phiSB\gtrsim 1$ examined in Ref.~\cite{khodjasteh2007performance}, while distinguishing between the computational and memory settings. Here, the $\gtrsim$ symbol specifies that $\phiB>\phiSB$ but both are similar in magnitude so that both can be considered to be of the same order in a perturbative expansion. 

We focus on the computational setting for now. Following the analysis of Ref.~\cite{khodjasteh2007performance}, the effective Hamiltonian for $\CDDn$ can be written as
\begin{align}\label{eq:cdd-generator-est}
\Omega_{\CDDn} &=  I \otimes 4^n \tau B_I \nonumber \\
&+ X \! \otimes\! (-\upi)^{n} 2^{n(n+1)} \tau^{n+1}\!\ad_{B_I}^{n}\!(B_X)\\ 
&+ Y \! \otimes\! (-\upi)^{n} 2^{n^2}\!\!\tau^{n+1}\ad_{B_I}^{n-\!1}\!( [B_I,B_Y] - \upi \{B_X,B_Z\} )\nonumber\\
&+\text{(\,higher order terms\,)}. \nonumber
\end{align} 
We confirm this leading-order behavior in Appendix \ref{app:CDD}, an analysis needed also for our $B_I=0$ discussion below. $\CDDn$ thus indeed achieves $n$th-order decoupling in that the lowest-order system-bath interaction term is $(n+1)$th order in $B_I$ and/or $B_\alpha$; lower-order terms have been eliminated by the DD sequence. 
The pure-bath part, whose size can be approximated  by
\begin{equation}
\Phi_{\mathrm{B},n} \simeq 4^n\phiB, 
\end{equation}
is of the first order. 
The remnant interaction part can be bounded in a straightforward manner using the leading-order terms in Eq.~\eqref{eq:cdd-generator-est}, giving the error phase
\begin{equation}\label{eq:CDDerrph-ub}
\Phi_{\mathrm{SB},n}\lesssim 2^{(n+1)^2}\phiB^n\phiSB.
\end{equation}

\begin{figure}
\includegraphics[width=0.9\columnwidth]{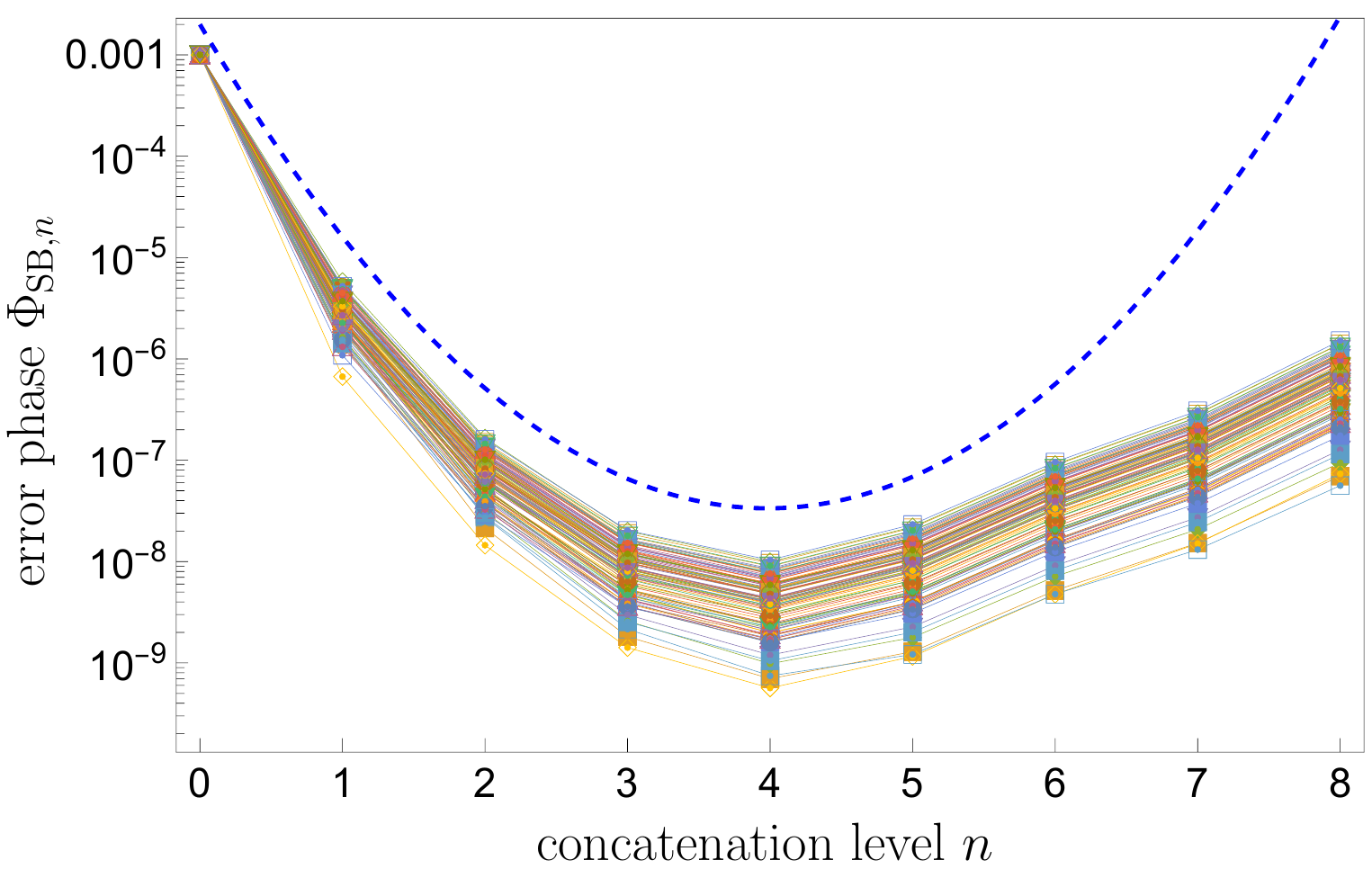}
\caption{The error phase $\Phi_{\rSB,n}$ as a function of the CDD concatenation level. 
The thin colored lines are obtained from random samples of the Hamiltonian $H$ satisfying $\phiB=\phiSB=0.001$. The blue dashed line is our theoretical upper bound for the error phase. The  maximal concatenation level occurs at 4 for these parameter values.}
\label{fig:estimator-size}
\end{figure}

That $\phiB$ enters the error phase should again be of no surprise. As in PDD, $\phiB$ determines how quickly the noise seen by the system evolves, and hence affects the efficacy of DD designed to eliminate noise that remains unchanged for the full DD sequence. $\phiB$ has to be small for DD to work well. 
What is perhaps more surprising is the $2^{(n+1)^2}$ factor in $\Phi_{\mathrm{SB},n}$. The origin of this factor lies in the exponentially increasing temporal length of the CDD pulse sequence as $n$ increases, for fixed $\tau_0\equiv \tau$ (implicit in the $\phiB$ and $\phiSB$ quantities). This means that the error phase eventually increases for large enough $n$, as the exponentially decreasing $\phiB^n$ factor is eventually overcome by the super-exponentially increasing $2^{(n+1)^2}$ factor. This tells us, as anticipated in the introductory paragraphs to this section, that \emph{there is no accuracy threshold}: There are no nonzero values for $\phiB$ and $\phiSB$ below which $\Phi_{\mathrm{SB},n+1}\leq\Phi_{\mathrm{SB},n}$ for \emph{all} $n$. Instead, there is a maximal useful level of CDD, beyond which further concatenation actually increases the noise seen by the system. Fig.~\ref{fig:estimator-size} plots this situation of fixed $\tau_0$, for increasing concatenation level $n$. We observe the initial decrease of $\Phi_{\mathrm{SB},n}$ as $n$ increases, but this turns around eventually (at $n=4$ for the plotted situation). 

Of course, the above discussion is based on the upper bound for the error phase. We can only conclude that there is no threshold level of noise below which the \emph{upper bound} on error phase decreases as $n$ increases. A more careful discussion of the threshold should look at the original $\Omega_{\CDDn}$ expression, rather than a bound. For that, we examine the actual map 
$\Omega_{\CDD{n+1}}\equiv \cD(\Omega_{\CDD{n}})$ that recursively gives the evolution as the CDD level increases, leading to Eq.~\eqref{eq:cdd-generator-est}. Stated in terms of the bath operators and neglecting higher-order terms, the map $\cD$ is given by (see App.~\ref{app:CDD}), for $n\geq 0$,
\begin{align}
 \tau B_{n+1,I} &= 4  \tau B_{n,I}=4^{n+1} \tau B_I,\label{eq:cdd-update}\\
 \tau B_{n+1,X} & = -4\upi [ \tau B_{n,I}, \tau B_{n,X} ],\nonumber\\
 \tau B_{n+1,Y} & = -2\upi  [ \tau B_{n,I}, \tau B_{n,X} ],\nonumber\\
 \tau  B_{n+1,Z} &=0,\nonumber
\end{align}
where $B_{n,\alpha}$, for $\alpha=I,X,Y,Z$, refers to the bath operators in $\Omega_{\CDDn}$ associated with $\sigma_\alpha$ on the system for $\CDDn$: $\Omega_{\CDDn}=\sum_{\alpha}\sigma_\alpha\otimes\tau B_{n,\alpha}$, with $\Omega_{\mathsf{CDD}0}\equiv \Omega$, and $B_{0,\alpha}=B_\alpha$; the factor of $\tau$ that accompanies every $B$ operator makes it a dimensionless quantity. Since $B_{n,Z}$=0 at this order of approximation for $n\geq 1$, for the norm of the interaction part of $\Omega_{\CDDn}$ to decrease as $n$ increases, it suffices to require $\tau B_{n,X}$ and $\tau B_{n,Y}$ to each decrease in size as $n$ grows. This happens if the map $-4\upi [\tau B_{n,I},\,\cdot\,]$ is contractive, satisfied if $8\Vert\tau B_{n,I}\Vert<1$. This gives a maximal concatenation level,
\begin{equation}\label{eq:cdd-max-level}
n_{\max} = \lceil -\log_4(\phiB)-\tfrac{3}{2} \rceil,
\end{equation}
where $\lceil~\cdot~\rceil$ denotes the ceiling function. At $n=n_{\max}$, the update rules Eq.~\eqref{eq:cdd-update}] no longer give a contractive map, and the error phase can grow upon further concatenation.
For $\phiB=0.001$, this sufficient condition successfully predicts the $n_{\max}=4$ numerically observed in Fig.~\ref{fig:estimator-size}. The maximal concatenation level depends only  on the norm of $\tau B_I$ here, but this need not hold in all settings.
It is worth mentioning that in Refs.~\cite{zhang2007dynamical,*zhang2008longtime}, the authors find through numerical investigations that an optimal decoupling strategy is $\CDD2$, when compared with other schemes like PDD, Symmetric DD (a time-symmetrized version of PDD), and $\CDD4$. The existence of such an optimal strategy is consistent with our observation of a maximal concatenation level here.

The existence of a maxima concatenation level, beyond which further concatenation increases---rather than decreases---the noise strength, does not technically contradict the statement that 
higher-order decoupling is achieved by higher-level CDD. Using decoupling order to quantify the degree of noise removal requires, in the first place, the convergence of the Magnus series, so that the $(n+1)$-th and higher-order terms in the series are small corrections to the first $n$ terms. However, if the DD scheme is designed such that the total time for the sequence grows exponentially with $n$, as is the case for CDD if $\tau_0$ is fixed as $n$ increases, eventually, we violate the convergence criterion and the decoupling order stops being a reasonable indicator of successful noise removal. 

That the total sequence time grows exponentially with $n$ also provides the intuition to the non-existence of an accuracy threshold for CDD, and that CDD fails after some maximal level, even in the ideal (perfect pulses) case. As $n$ grows, the sequence time grows as $4^n\tau$ in this computational setting where the per-pulse time interval is some fixed value $\tau$ independent of $n$. Recalling that DD works to average away noise that remains constant for the full sequence time, as $n$ grows, the sequence time eventually becomes long compared to the time-scale of evolution of the noise, reducing the efficacy of the DD averaging. Eventually, the additional concatenation adds only to the total sequence time, without adding much noise-removal power, and we reach the maximal useful concatenation level beyond which further concatenation makes the total (over the full sequence time) noise worse.

To further confirm this intuition, it is useful to momentarily consider the memory setting appropriate for a a different physical situation: to have fixed $\tau_n= T\equiv \tau$, so that the CDD sequence takes the same amount of time, regardless of $n$. This requires per-pulse time interval $\tau_0=\frac{\tau}{4^n}$ for each $n$, so that the physical pulses are applied at shorter and shorter time intervals as the concatenation level increases, up to some practical limit in the pulse rate. In this case, the evolution can be written as
\begin{equation}
 \Omega_{n+1} = \cD(\Omega_{n}/4),
\end{equation}
so that the recursive rules of Eq.~\eqref{eq:cdd-update} become
\begin{align}\label{eq:cdd-update2}
\tau B_{n+1,0} &=\tau B_{n,0},\\
 \tau B_{n+1,1} & = -\tfrac{1}{4}\tau^2\upi [B_{n,0}, B_{n,1} ],\nonumber\\
 \tau B_{n+1,2} & = -\tfrac{1}{8}\upi \tau^2[B_{n,0}, B_{n,1} ],\nonumber\\
 \tau  B_{n+1,3} &=0.\nonumber
\end{align}
This gives
\begin{align}\label{eq:cdd-generator-est2}
&\Omega_{n} 
\simeq  \sigma_0\!\otimes\!\tau B_I \notag \\
&+ \sigma_1 \!\otimes\! (-\upi)^{n} 2^{-n(n+1)} \tau^{n+1}\ad_{B_I}^{n}(B_X)\\ 
& +\sigma_2 \!\otimes\! (-\upi)^{n} 2^{-n(n+2)} \tau^{n+1}\ad_{B_I}^{n-\!1}([B_I,B_Y] - \upi \{B_X,B_Z\}\!). \notag
\end{align} 
$\Phi_{\mathrm{B},n}= \phiB$ thus remains unchanged, while the error phase is reduced to
\begin{equation}
 \Phi_{\mathrm{SB},n} \lesssim \,2^{-n^2} \phiB^{n}\,\phiSB,
\end{equation}
where we have dropped the sub-leading-order term. For reasonably bounded noise ($\phiB<2$), the condition $\Phi_{\mathrm{SB},n+1}<\Phi_{\mathrm{SB},n}$ automatically holds for all $n\geq 0$. The error phase is super-exponentially decreasing in $n$, suggesting---in sharp contrast to the computational setting---that the noise can be arbitrarily suppressed by increasing the CDD level, limited only by practical constraints on how small $\tau_0$ can be in the experiment.

\subsubsection{\texorpdfstring{$\phiB/\phiSB\ll 1$}{\straightphi B/\straightphi SB << 1}}

Next, we consider the situation where $B_I$ is negligible compared with $B_\alpha$ so that we effectively set $B_I=0$ (and hence $\phiB$) to the approximation order considered below. As mentioned earlier, one expects CDD to perform better in this case: DD is intended to remove slow---compared with the gate times---noise, the limiting situation being one where $B_I=0$ so that the noise in fact does not evolve by itself in the absence of the system. As we will see below, this $B_I=0$ limit, however, does not mean that an accuracy threshold for CDD exists. The behavior of CDD as the concatenation level increases in fact eventually resembles that of CDD with $B_I\neq 0$. This can be understood intuitively once we recognize---as we will see below---that the CDD pulses cause a back-action---through the system-bath interaction---on the bath itself, such that the effective $B_I$ after adding the DD pulses becomes nonzero, and eventually we get back the situation of the previous subsection.

 When $B_I=0$ (or of higher order in our approximation compared with $B_\alpha$),  the terms in Eq.~\eqref{eq:cdd-generator-est} for the effective Hamiltonian for $\CDDn$ vanish, except when $n=1$ (the PDD case; there, we retain the $\sigma_2\otimes\{B_X,B_Z\}$ term). That expression is hence no longer useful beyond PDD, as is the case of the recurrence map Eq.~\eqref{eq:cdd-update}. Both expressions came from considering only the first and second-order Magnus terms for every $n$ (see App.~\ref{app:CDD}). 
 
For a non-vanishing contribution for $n>1$, we first consider the third-order term for $n=1$, the PDD case; we will see why, momentarily. Straightforward algebra gives, for negligible $B_I$,
\begin{align}
\Omega_\PDD^{(3)}&=\tfrac{2}{3}\tau^3{\left[\upi I\otimes{\bigl([B_Z,\{B_X,B_Y\}]-[B_Y,\{B_X,B_Z\}]\bigr)}\right.}\nonumber\\
&~\qquad{\left.+3\sigma_z\otimes\{B_X,\{B_X,B_Z\}\}\right]}.
\end{align}
For PDD, this third-order correction is sub-leading order, compared with the non-vanishing second-order $\Omega_\PDD^{(2)}$, and can be neglected in the PDD analysis. However, we see that this third-order term gives rise to a nonzero $B_{1,0}$, the bath operator associated with the identity on the system in $\Omega_\PDD$. This term arises solely from the $\HSB$ bath operators $B_X, B_Y$, and $B_Z$, and can be thought of as a back-action on the bath due to its interaction with the system.

Given the recursive structure of CDD, this means that, while at level $n=1$, the PDD analysis has $B_I=B_{0,0}=0$, at the next level $n=2$, we no longer have this no-pure-bath-evolution situation, since $B_{1,0}\neq 0$, albeit of a higher order. If we begin our recursion map $\cD$ at $n=2$, rather than $n=1$, using $\Omega_\PDD$ as the base dimensionless Hamiltonian in place of the physical $\tau H$, we are in essence back in the previous case where the pure-bath evolution is no longer vanishing. The former form of $\cD$ as specified in Eq.~\eqref{eq:cdd-update} thus holds for $n\geq 2$, with now the base case as $n=1$, not $n=0$, with the $B_{1,\alpha}$s given by
\begin{align}
\tau B_{1,0}&\equiv \upi\tfrac{2}{3}\tau ^3{\bigl([B_Z,\!\{B_X,B_Y\!\}]\!-\![B_Y\!,\!\{B_X,B_Z\}]\bigr)}\!+\!O(\phiSB^4)\nonumber\\
\tau B_{1,1}&\equiv 0+O(\phiSB^4)\\
\tau B_{1,2}&\equiv -2\tau ^2\{B_X,B_Z\}+O(\phiSB^4)\nonumber\\
\tau B_{1,3}&\equiv 2\tau^3 \{B_X,\{B_X,B_Z\}\}+O(\phiSB^4).\nonumber
\end{align}
The dimensionless Hamiltonian for $\CDD{n}$ is then estimated for $n\geq 2$, following Eq.~\eqref{eq:cdd-generator-est}, as
\begin{align}
\Omega_{\CDD{n}}&\simeq \sigma_0 \otimes 4^{n-1}\tau B_{1,0} \\
&\quad+ \sigma_2\! \otimes\! (-\upi)^{n-1} 2^{(n-1)^2}\tau^n \ad_{B_{1,0}}^{n-\!1}\!(B_{1,2}). \nonumber
\end{align}
Here, we have kept only the lowest-order terms that contribute to $H_{\rB,n}$ and $H_{\rSB,n}$, keeping in mind the various $\tau B_{1,\alpha}$s are of different orders of magnitude: $\tau B_{1,0}$ and $\tau B_{1,3}\sim \phiB^3$, $\tau B_{1,1}\sim O(\phiSB^4)$, and $\tau B_{1,2}\sim O(\phiSB^2)$.

\begin{figure}[ht!]
    \includegraphics[width=0.9\columnwidth]{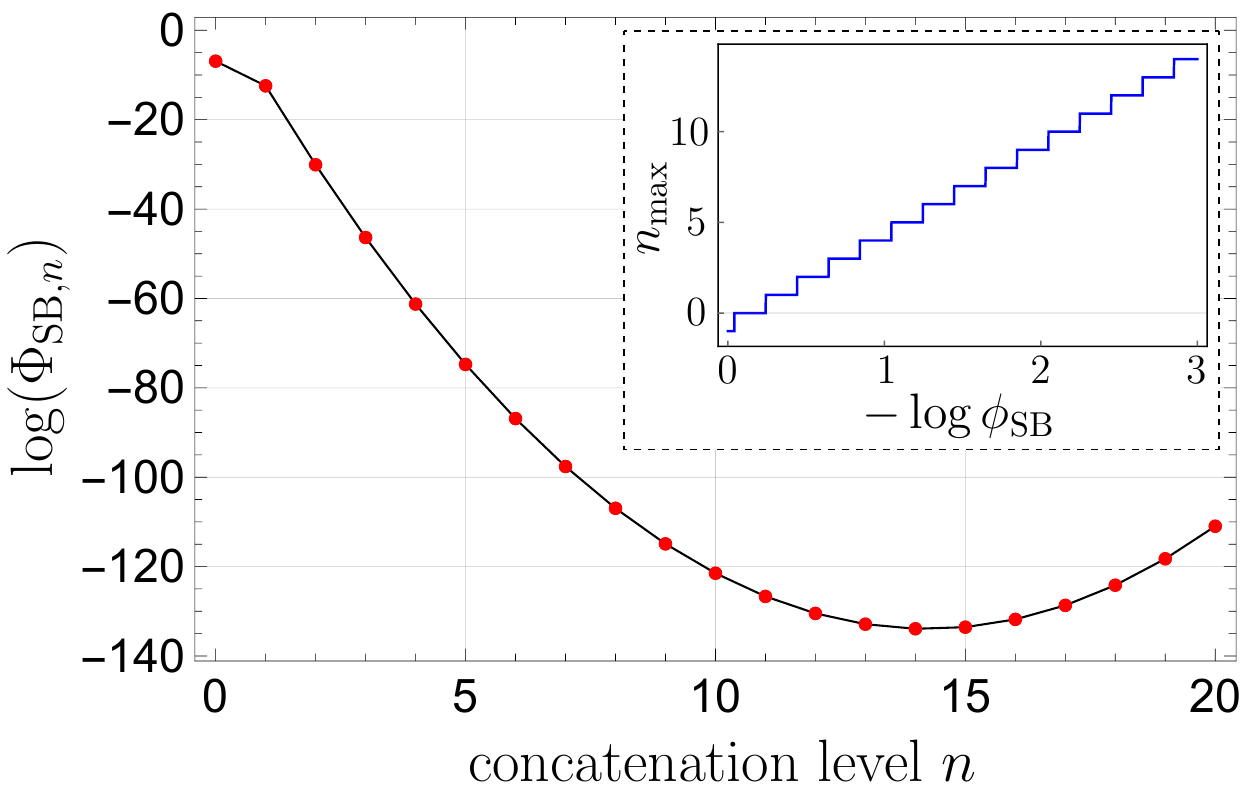}
    \caption{ 
    The error phase as a function of the concatenation level using the theoretical upper bound of Eq.~\eqref{eq:zeroB0-upperbound}, for $\phiB=0$ and $\phiSB=0.001$. The maximal concatenation level occurs at $n=14$ for this particular configuration. The inset shows the maximal concatenation level as a function of the error phase, as predicted by \eqref{eq:n-zeroB0}. Non-positive $n$ values here indicate that even PDD (n=1) does not offer any benefit.
    }
    \label{fig:zeroB0}
\end{figure}

In this case, the error phase for level-$n$ CDD can be bounded as
\begin{align}\label{eq:zeroB0-upperbound}
\Phi_{\rSB,n}&\lesssim 2^{n(n-1)}\Vert\tau  B_{1,0}\Vert^{n-1}\Vert\tau  B_{1,2}\Vert\nonumber\\
&\lesssim\frac{2^{n^2+3n-2}}{3^{n-1}}\phiSB^{3n-1}={\left(\tfrac{8}{3}\right)}^{n-1}2^{n^2+1}\phiSB^{3n-1},
\end{align}
noting that $\Vert\tau  B_{1,0}\Vert\lesssim \frac{16}{3}\phiSB^3$ and $\Vert\tau  B_{1,2}\Vert\lesssim 4\phiSB^2$.
Comparing this with the error phase bound of $\Phi_{\rSB,n}\sim \phiB^n\phiSB$, the current situation of $B_I=0$ gives a faster suppression of $\phiSB^{3n}$ as $n$ increases. The numerical prefactor still grows exponentially with $n$, however, so we still expect the accuracy threshold for CDD to vanish even in this case. Figure \ref{fig:zeroB0} plots the error-phase bound for this $B_I=0$ case for $\phiSB=0.001$. This shows the same qualitative behavior as in Fig.~\ref{fig:estimator-size}, where $B_I\neq 0$, though the maximal $n$ value is larger in this $B_I=0$ situation.

A similar analysis as in the $\phi_\rB/\phi_\rSB\gtrsim 1$ regime leads to an analogous maximal concatenation level for CDD to offer benefit in this $B_I=0$ setting,
\begin{equation}\label{eq:n-zeroB0}
n_{\max}=\lceil-3\log_4(\phiSB)+\log_43-\tfrac{5}{2}\rceil,
\end{equation}
where we have used the fact that $B_{n+1,0}=4^nB_{1,0}$ for $n\geq 2$. For $\phiSB=0.001$, this gives $n_{\max}=14$, which 
agrees with the maximal $n$ level observed in Fig.~\ref{fig:zeroB0}.
We note that, compared with the $\phi_\rB/\phi_\rSB\gtrsim 1$ regime of the previous subsection, this $\phi_\rB\ll1 $ situation presents much more favorable conditions for the performance of CDD, both in terms of the smallest achievable error phase and the maximal concatenation level (c.f.\ Figs.~\ref{fig:estimator-size} and \ref{fig:zeroB0}).
The inset to Fig.~\ref{fig:zeroB0} shows the maximal useful CDD level as a function of $\phiSB$, using Eq.~\eqref{eq:n-zeroB0}. We see that DD becomes useful when $\phiSB\lesssim 10^{-0.25}\simeq 0.56$, an improvement compared with the prediction of the generic Condition \eqref{eq:PDDCond} (with $\phiB=0$, giving $\phiSB\lesssim 0.25$) based on a loose bound. These results support the intuition that DD performs better in the $\phiB\ll\phi_\rSB$ regime, although the quantitative behavior, in particular the existence of a maximal concatenation level, is similar in both.

\subsection{Noisy CDD}
Next, we discuss the CDD performance when the DD gates themselves are noisy. In particular, we are interested in the strength of the combined noise by studying how the error phase upper bound will evolve with the CDD concatenation level. Since we want to understand the limitations of CDD, we focus only on the $\phi_\rB/\phi_\rSB\gtrsim 1$ regime, the scenario with poorer CDD performance, as explained earlier.

With imperfect control, the iterative map $\cD$ of the ideal situation gets modified by the noise, and we write $\widetilde\Omega_{\CDD{n+1}}=\widetilde\cD(\widetilde\Omega_{\CDD{n}})$ for $n\geq 0$, where, as before, a tilde indicates the noisy version of the quantity, and we have $\widetilde{\Omega}_{\CDD1}\equiv\widetilde{\Omega}_\PDD$ and $\widetilde{\Omega}_{\CDD0}\equiv \Omega$. The use of the same noisy map $\widetilde\cD$ stems from the fact that the same physical gates---with the same noise---are used to implement CDD at every level. The error phase, namely, the norm of the interaction part of each $\widetilde\Omega_{\CDD{n}}$, is denoted as $\widetilde\Phi_{\rSB,n}$. 

The accuracy threshold condition under noisy control is given by
\begin{equation}\label{eq:Phi}
\widetilde\Phi_{\rSB,n+1}\leq \widetilde\Phi_{\rSB,n},
\end{equation}
for every $n$. From our earlier discussion, we understand that there is no accuracy threshold even with perfect control, but it is still meaningful to study how noise can adversely impact the error phase at each level. For each $n$, the recursive nature of CDD allows us to view Eq.~\eqref{eq:Phi} as the breakeven condition for noisy PDD, with $\Omega$ replaced by $\widetilde\Omega_{\CDD{n-1}}$. We thus simply need to understand how $\widetilde\Phi_{\rSB,n}$ and $\widetilde\Phi_{\rB,n}$ evolve as $n$ increases, and make use of the PDD breakeven conditions discussed in Sec.~\ref{sec:PDD}.

We consider the computational setting, with only gate-control noise and, for simplicity, instantaneous pulses ($\tau_P=0$). The noisy $X$ and $Z$ gates are written as
\begin{equation}
\widetilde X=X\upe^{-\upi\Gamma_X} \quad\textrm{and}\quad\widetilde Z=Z\upe^{-\upi\Gamma_Z},
\end{equation}
where we recall that $\Gamma_X$ and $\Gamma_Z$ describe the gate-control noise, and are generally operators on both the system and the bath. We write $\Gamma_{X(Z)}\equiv \sum_{i=1}^3\sigma_i\otimes\Gamma_{X(Z),i}$ where $\Gamma_{X(Z),i}$ acts only on the bath. $\Gamma_{X(Z)}$, as previously noted, can be taken to have no pure-bath term, and $\Vert\Gamma_{X(Z)}\Vert\leq \eta$. Our earlier calculation [Eq.~\eqref{eq:noisyOpdd2}] for noisy PDD gives, as the first-order Magnus term,
\begin{equation}\label{eq:CDDO1}
\widetilde{\Omega}_\mathsf{PDD}^{(1)}\simeq(4\tau)I\otimes B_I+2Y\otimes (\Gamma_{X,2}-\Gamma_{Z,2}).
\end{equation}
From this, we know that the pure-bath term approximately (ignoring higher-order corrections) quadruples in size with every concatenation level,
\begin{equation}\label{eq:PhiBn}
\widetilde\Phi_{\rB,n}\simeq 4\widetilde\Phi_{\rB,n-1}=4^n\phiB.
\end{equation}
For the interaction part, the first-order Magnus expression gives a term with norm bounded by $4\eta$. Since $\CDD{n}$ is just PDD on $\CDD{(n-1)}$ with the same noisy gates, this term will occur unchanged at all concatenation levels, giving (again, ignoring higher-order corrections)
$\widetilde\Phi_{\rSB,n}\lesssim 4\eta$,
an $n$-independent bound. This is the appropriate estimate of the error phase if $\eta$ is dominant over the $\phiSB$ and $\phiB$ terms that arise in the second-order Magnus term. In the opposite limit, where $\eta\ll \phiSB,\phiB$, we should recover the ideal PDD behavior, where the first-order Magnus term has no interaction piece, and the leading order correction is the second-order Magnus term [Eq.~\eqref{eq:PDDMag2}] for ideal PDD. In this case, we estimate $\widetilde\Phi_{\rSB,n}$ as in Eq.~\eqref{eq:Opdd2}, with $\phiSB$ and $\phiB$ replaced by $\widetilde\Phi_{\rSB,n-1}$ and $\widetilde\Phi_{\rB,n-1}$, respectively, the noise parameters from $\CDD{(n-1)}$ . Putting the two limits together, we can estimate the error phase as,
\begin{equation}\label{eq:PhiSBn}
\widetilde\Phi_{\rSB,n}\lesssim 12\widetilde\Phi_{\rB,n-1}\widetilde\Phi_{\rSB,n-1}+4\widetilde\Phi_{\rSB,n-1}^2+4\eta,
\end{equation}
now valid for all values of $\eta$. Equations \eqref{eq:PhiBn} and \eqref{eq:PhiSBn} give a pair of recurrence relations that tells us how the noise parameters evolve as $n$ increases.

\begin{figure}[htb]
 \includegraphics[width=0.95\columnwidth]{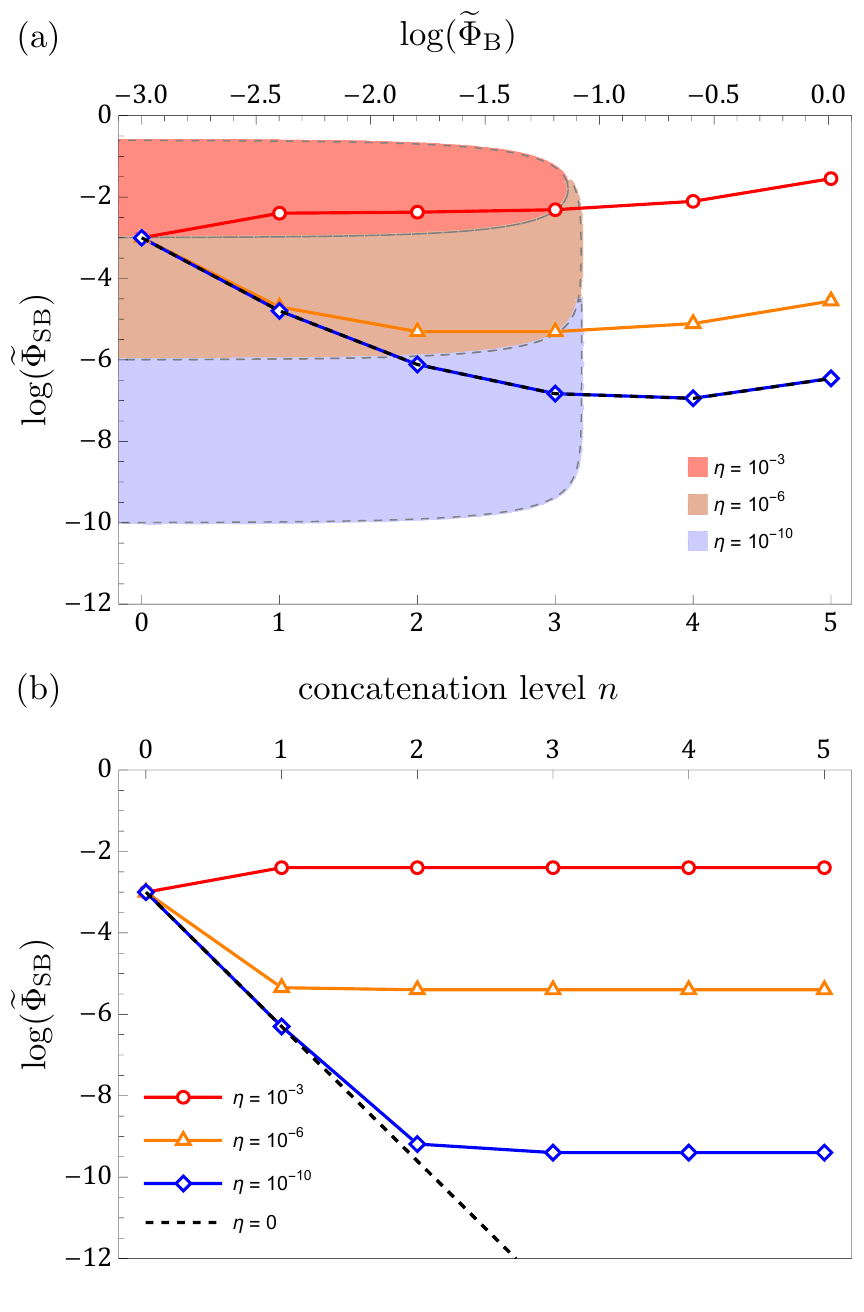}
 \caption{The evolution of the $(\widetilde\Phi_{\rB},\widetilde\Phi_{\rSB})$ pair with increasing concatenation level, for (a) the computational setting, and (b) the memory setting. In both plots, the three colored lines, from top to bottom, are for the overall pulse errors of $\eta=10^{-3}, 10^{-6}$ and $10^{-10}$ respectively,while the black dashed line gives the noiseless limit (in (a) it is nearly coincident with the $\eta = 10^{-10}$ line).
 The leftmost marker for each line is the $n=0$ (i.e., no DD) point, and we have $n=1,2,3,\ldots$ for the subsequent markers as we move rightwards on the line. The shaded regions in (a) are the PDD ($n=1)$ breakeven regions for the different values of $\eta$ in the computational setting.}
 \label{fig:cddFig}
\end{figure}

Intuitively, we can understand the origin of the maximal useful CDD level of concatenation by combining the recurrence relations with the PDD breakeven condition.
From our understanding of PDD in Sec.~\ref{sec:PDD}, we know that the breakeven condition for PDD identifies a bounded region for $\phiB$ and $\phiSB$.  Yet, we see here that $\widetilde\Phi_{\rB,n}$---which enters the bound for $\widetilde\Phi_{\rSB,n}$---grows without bound, indicating that there is no accuracy threshold, i.e., no values of the noise parameters for which Condition~Eq.~\eqref{eq:Phi} is satisfied for all $n$. Such is to be expected, given our conclusion for ideal CDD. We can, moreover, determine the maximal useful CDD level of concatenation from this: The maximal level corresponds to the $n$ such that the $(\widetilde\Phi_{\rB,n}, \widetilde\Phi_{\rSB,n})$ pair just crosses the boundary determined by the PDD breakeven condition.
We illustrate this point in Fig.~\ref{fig:cddFig}(a), 
where three different systems with control noise levels $\eta=10^{-3},10^{-6}$, and 
$10^{-10}$ are explored, indicated in red, orange, and blue shades, respectively.
Figure \ref{fig:cddFig}(a) plots the regions bounded by the breakeven conditions for those systems, and simulates the evolution of $(\widetilde\Phi_{\rB},\widetilde\Phi_{\rSB})$ for the effective Hamiltonian as one increases the concatenation level, for the same starting values of $\phiB=\phiSB=0.001$.  Since $\log(\widetilde\Phi_{\rB})$ is linear in $n$ (see \Eqref{eq:PhiBn}), the concatenation level can be alternatively read off from the horizontal axis.
For small enough $\eta$, we observe a similar behavior as in the ideal CDD case: $\Phi_{\rSB,n}$ first decreases then increases as $n$ grows, and we can identify a maximal $n$ beyond which scaling up CDD gives no further benefit.  That maximal $n$ value is the very last level where its previous level still lies within the breakeven region (not counting the boundary). For $\eta=10^{-3}$, we see that even $n=1$ is of no use: The error phase $\Phi_{\rSB,n}$ for every $n$ is larger than if no DD is used, as the bare Hamiltonian lies on the breakeven boundary to begin with.

Again, we can examine the memory setting for comparison. In that case, $\widetilde\Phi_{\rB,n}\simeq \phiB$ is approximately constant across the concatenation levels, while the error phase, following a similar logic as above, updates according to the rule,
\begin{equation}
\widetilde\Phi_{\rSB,n}\lesssim \frac{1}{2}\phiB\widetilde\Phi_{\rSB,n-1}+4\eta.
\end{equation}
Assuming the upper bound provides a good estimate of the actual quantity, we solve this recursive relation to obtain,
\begin{equation}\label{eq:CDDeta}
\widetilde\Phi_{\rSB,n} \simeq (\tfrac{1}{2}\phiB)^n \phiSB + \frac{4\eta}{1-\tfrac{1}{2}\phiB},
\end{equation}
keeping terms up to order $n+1$ in small quantities. As long as $\phiB<2$, $\widetilde\Phi_{\rSB,n}$ decreases as $n$ increases. However,  unless the control pulses are perfect, there will always be an $n$-independent remnant---the second term in Eq.~\eqref{eq:CDDeta} proportional to $\eta$---that will limit how small $\widetilde\Phi_{\rSB,n}$ can be, not to mention the eventual limit in $n$ when $\tau$($=T/4^n)$ for this memory setting gets too short to be feasible. 
As illustrated in Fig.~\ref{fig:cddFig}(b),  the error phase reduces and plateaus to the same order as $\eta$ as one increases the concatenation level.  Thus, in this case of noisy pulses, there is a limit in the error suppression capability of CDD, determined entirely by the size of the gate control noise. Hence, arbitrary noise suppression with CDD is not possible with imperfect control even in the memory setting.
Note that this conclusion does not rely on us equating $\widetilde\Phi_{\rSB,n}$ with its upper bound above, but simply results from the linear-in-$\eta$ term in Eq.~\eqref{eq:CDDO1}.

\section{Conclusions}\label{sec:Conc}

\begin{table*}[ht]
\caption{\label{tab:summary}Summary of results.}
\renewcommand{\arraystretch}{1.5}
\begin{tabular}{c|c|c|c}
\hline\hline
Assumptions&Setting&Condition&Eq(s).~in text\\
\hline\hline
\multicolumn{4}{c}{PDD, breakeven conditions}\\
\hline\hline
ideal&computational&$12\phiB+4\phiSB\leq1$&\eqref{eq:PDDCond}\\
\hline
noisy, general&computational&$ (8/\pi)\, \phi_\mathrm{SB}\,\delta +4\eta\leq \phi_\mathrm{SB}$&\eqref{eq:BE2}\\
\hline
noisy, $\eta=0$&computational&$\tau_P/\tau\leq 8/\pi \simeq 0.4 $&\eqref{eq:BE2eta0}\\
\hline
noisy, $\tau_P=0$&computational&$\eta\leq \phi_\mathrm{SB}/4$ &\eqref{eq:BE2tauP0}\\
\hline
 \begin{tabular}[c]{@{}c@{}}unitary control\\ errors only, $\tau_P=0$\end{tabular}&computational&
 $\begin{aligned}
  \eta&\leq \sqrt{\phi_\rSB/2}{\left[1-4(3\phi_\rB-\phi_\rSB)\right]}^{-\tfrac{1}{2}}-2\phi_\rSB\\
  &\approx \sqrt{\phi_\rSB/2} \  (\text{when }\phi_\rB, \phi_\rSB\ll 1)
 \end{aligned}$
 &\eqref{eq:unitaryBE}\\
\hline\hline
\multicolumn{4}{c}{CDD, maximal concatenation level}\\
\hline\hline
ideal, $\phiB\gtrsim \phiSB$&computational& $\lceil -\log_4(\phiB)-3/2 \rceil$ &\eqref{eq:cdd-max-level}\\
\hline
ideal, $\phiB \gtrsim \phiSB$ &memory& no maximal $n$, with $\Phi_\mathrm{SB}\to 0$ &-\\
\hline
ideal, $\phiB=0$&computational& $\lceil -3\log_4(\phiSB)+\log_43-5/2 \rceil$ &\eqref{eq:n-zeroB0}\\
\hline
noisy, $\tau_P=0$&computational&recurrence relations for maximal~$n$&\eqref{eq:PhiBn} and \eqref{eq:PhiSBn}\\
\hline
noisy, $\tau_P=0$&memory& no maximal $n$, with $\widetilde\Phi_{\rSB,n}\rightarrow 4\eta/(1-\phiB/2)$&\eqref{eq:CDDeta} \\
\hline\hline
\end{tabular}
\end{table*}

In this work, we studied the question of the fault tolerance of DD, namely, the efficacy of DD schemes in the presence of imperfections in the very pulses that carry out the DD operations. Our results are summarized in Table \ref{tab:summary}. We examined first the breakeven conditions on the noise parameters for single-level PDD, and then we analyzed the performance of CDD. We saw that, in the computational setting, of primary relevance today, there is generally a limit to the CDD concatenation level, beyond which further concatenation offers no added benefit. This is in contrast to the memory setting, where unlimited error suppression is possible in some cases: Here, we saw this for the case of ideal CDD, while the case of finite-width pulses gave similar conclusions in Ref.~\cite{khodjasteh2007performance}. In the case where gate control noise dominates, CDD faces a limit even in the memory setting, determined by the strength of the gate control noise.

That CDD faces a limit eventually is simply a reflection of the fact that the benefit, namely the error suppression capability, of CDD fails to grow sufficiently fast to compensate for the added cost, namely the additional gate control noise, from the increased number of pulses needed for larger $n$. It is easy to understand how this arises in the computational setting: Even in the case of ideal pulses, the exponential lengthening of the pulse sequence time as the CDD level increases means a relative increase in the noise evolution rate [see Eq.~\eqref{eq:PhiBn}]---whether the noise changes quickly or slowly is relative to the pulse sequence time---until eventually, for large enough $n$, the noise changes much faster than the pulse sequence time, and DD stops being effective. This is exacerbated by imperfections in the pulses, with the limiting $n$ reached earlier than in the ideal case; see Fig.~\ref{fig:cddFig}(a) as an example. In contrast, the memory setting sees no such relative increase in noise evolution rate as, in this case, all $\CDDn$ sequences, for any $n$, are assumed to take the same total time.

That gate control noise sets a limit to the error suppression capability of CDD can also be understood intuitively. Any gate-dependent errors is equivalent to noise that changes during the pulse sequence and hence cannot be effectively averaged away by the DD sequence. This is in contrast to any gate-independent noise, which includes noise that arises from the finite-width pulses due to the always-on and time-independent $\HSB$ during the pulse application. Such static noise---referred to as ``systematic errors" in Refs.~\cite{khodjasteh2007performance, khodjasteh2005fault}---can be removed by the DD sequence. Note that Refs.~\cite{khodjasteh2007performance,khodjasteh2005fault} hinted at some robustness of CDD against what they termed ``random errors", as opposed to systematic errors. Our gate-dependent gate control noises are closer to the random errors mentioned in Refs.~\cite{khodjasteh2007performance,khodjasteh2005fault}, but they are not quite random enough for truly randomized averaging effects, hinted at in Ref.~\cite{khodjasteh2007performance}, to kick in. Yet, such kinds of gate-dependent errors are generic in experiments today, with the same noise statistics manifesting each time the same gate is applied.

While we chose PDD and CDD as our focus here, the insights gained from them apply to arbitrary scalable DD schemes. As long as the DD pulse sequence gets longer as one scales up, the above intuition that the noise evolution rate increases relatively continues to hold true, and one again expects a limit in the efficacy of CDD.

Lastly, it is worthwhile to note that, even though we conclude here that there is a limit to the usefulness of CDD concatenation in the realm of realistic pulses, we should remember that CDD is anyway not meant as the final solution to noise in a quantum computing device. Instead, it is a lower-level means to remove slow noise, with higher-level methods like error correction taking over eventually to eliminate any remnant noise. The advantage of CDD, over that of error correction, is its much lower-resource requirements, and this remains a key consideration in near- to middle-term quantum devices. One can simply employ CDD to the maximum $n$ limit, weakening the noise as much as possible, before switching over to more expensive, but more powerful, quantum error correction.

\acknowledgments
We are grateful for discussions with Gu Yanwu, Jonas Tan, and Ryan Tiew, who did some of the preliminary investigations into the effects of noise in dynamical decoupling that eventually led to the work described here. This work is supported by the Ministry of Education, Singapore (through Grant No. MOE2018-T2-2-142).

\bibliographystyle{apsrev4-2}
\bibliography{dd-references}

\newpage

\appendix

\section{Relating the error phase and infidelity measures}\label{app:epInF}

Here, we relate the error phase (see Sec.~\ref{sec:PrelimEP}) and infidelity (see Sec.~\ref{sec:PrelimInFid}) measures. We consider the system and bath evolution for some time $T$, with effective dimensionless Hamiltonian $\Omega$ ($U(T,0)=\upe^{-\upi\Omega}$), using the appropriate $\Omega$ for the bare evolution without DD, or in the case with DD. Given our context of weak noise and slow evolution of the bath for good DD performance, we assume $\Omega$ to be small in norm and write down the system-bath evolution as a power series in $\Omega$ using the Baker-Hausdorff lemma \cite{rossmann2006lie}, so that we have
\begin{align}
\cN(\rhoS)&=\tr_\rB\bigl\{U(T,0)(\rhoS\otimes\rhoB)U(T,0)^\dagger\bigr\}\nonumber\\
\label{eq:powerSeries}&=\sum_{n=0}^\infty\frac{(-\upi)^n}{n!}\tr_\rB\bigl\{\ad_\Omega (\rhoS\otimes\rhoB)\bigr\},
\end{align}
where $\ad_\Omega$ is the map $\ad_\Omega(\cdot)\equiv [\Omega,\,\cdot\,]$. With this [see Eq.~\eqref{eq:trdisInFDef}], we can write
\begin{equation}
\infid(\cN,\psi)^2= \langle\psi|(\cI-\cN)(\psi)|\psi\rangle\equiv\sum_{n=1}^\infty f_n(\cN,\psi),
\end{equation}
for $f_n(\cN,\psi)\equiv-\frac{(-\upi)^n}{n!}\langle\psi|\trB\bigl\{\ad_\Omega^n(\psi\otimes\rhoB)\bigr\}|\psi\rangle=-\frac{(-\upi)^n}{n!}\tr\bigl\{\psi\,\ad_\Omega^n(\psi\otimes\rhoB)\bigr\}$, from the order-$n$ term in the power series Eq.~\eqref{eq:powerSeries}. This allows us to examine the infidelity measure as a series also in $\Omega$, with the leading orders giving an approximate expression.

To further evaluate the $f_n$s, we note two identities,
\begin{align}
\tr\{\psi\ad_A(\psi\otimes\rhoB)\}&=0,\label{eq:id1}\\
\textrm{and}\quad \trB\ad_{B}&=0,\label{eq:id2}
\end{align}
where $A$ is an arbitrary system-bath operator, while $B$ is a bath-only operator, i.e., acts nontrivially only on the bath. Eq.~\eqref{eq:id1} immediately tells us that the first-order term $f_1(\cN,\psi)$ always vanishes. For the second-order term, we first write $\Omega=\Omega_\rB+\Omega_\rSB$ with $\Omega_\rB\equiv I\otimes \trS\Omega$ as the bath-only part, and $\Omega_\rSB\equiv \Omega-\Omega_\rB$ is the system-bath term. Then, invoking the Jacobi identity, $\ad_A\ad_{A'}=\ad_{A'}\ad_A+\ad_{[A,A']}$, together with Eqs.~\eqref{eq:id1} and \eqref{eq:id2}, we find
\begin{equation}\label{eq:inf2}
f_2(\cN,\psi)=\tfrac{1}{2}\tr\{\psi\ad^2_{\Omega_\rSB}(\psi\otimes\rhoB)\}.
\end{equation}
$f_2(\cN,\psi)$ hence does not depend on $\Omega_\rB$ and scales as the square of the error phase $\Vert\Omega_\rSB\Vert^2$. In fact, from Eq.~\eqref{eq:inf2}, we can write down the following bound,
\begin{align}
f_2(\cN,\psi)&\leq\tfrac{1}{2}\Vert\rhoS\otimes I\Vert\,\Vert\ad_{\Omega_\rSB}^2(\rhoS\otimes\rhoB)\Vert_{\tr}\nonumber\\
&\leq 2\Vert\Omega_\rSB\Vert^2\,\Vert\rhoS\otimes\rhoB\Vert_{\tr}=2\Vert\Omega_\rSB\Vert^2.
\end{align}
Assuming $f_2(\cN,\psi)$ nonvanishing, we thus have that $\infid(\cN,\psi)^2\simeq f_2(\cN,\psi)\leq 2\Vert\Omega_\rSB\Vert^2$ for all $\cN$ and $\psi$, so that
\begin{equation}
0\leq \infid\lesssim \sqrt 2\Vert\Omega_\rSB\Vert.
\end{equation}

\section{Norm inequality}\label{app:NormIneq}
Here, we prove an inequality used in the main text,
\begin{equation}\label{eq:Bineq}
\max_{i\in\{1,2,3\}}\Vert B_i\Vert\leq \Bigl\Vert\sum_{i=1}^3\sigma_i\otimes B_i\Bigr\Vert,
\end{equation}
for $B_i$s Hermitian operators, $\sigma_i$ the usual Pauli operators, and $\Vert\cdot\Vert$ is the operator norm $\max_{x:\Vert x\Vert =1}|x^\transpose \cdot x|$. In the main text, we had used this inequality in the context where $B_i$s are the bath operators associated with $\sigma_i$s on the system in the interaction Hamiltonian $\HSB$.\\[1ex]
\noindent\underline{Proof}. We first prove a lemma: For $A_i$ Hermitian and $A_0$ positive semi-definite, we show that $\Vert I\otimes A_0+\sum_{i=1}^3\sigma_i\otimes A_i\Vert\geq\Vert A_0\Vert$. Let $M\equiv I\otimes A_0+\sum_{i=1}^3\sigma_i\otimes A_i$. It can be written in a block form, following the standard matrix representation of $\sigma_i$s, as
\begin{equation}
M={\left(\begin{array}{cc}
A_0+A_3&A_1-\upi A_2\\A_1+\upi A_2&A_0-A_3
\end{array}
\right)}
\end{equation}
Following the same block form, an arbitrary vector can be written as $(x,y)^\transpose $, with $x$ and $y$ vectors (columns) themselves. Let $x$ be the (unit-length) eigenvector corresponding to the greatest eigenvalue of $A_0$. Since $A_0\geq 0$, $\Vert A_0\Vert =x^\transpose A_0x$. We calculate the following,
\begin{align}
{\left(\!\begin{array}{cc}x^\transpose &0\end{array}\!\right)}M{\left(\!\begin{array}{c}x\\0\end{array}\!\right)}&=x^\transpose A_0x+x^\transpose A_3x=\Vert A_0\Vert+x^\transpose A_3x.
\end{align}
This means that $\Vert M\Vert\geq \Vert A_0\Vert+x^\transpose A_3x$. A similar argument with $(0~~x)^\transpose $ in place of $(x~~0)^\transpose $ gives $\Vert M\Vert\geq \Vert A_0\Vert-x^\transpose A_3x$. Combining the two inequalities gives the desired result: $\Vert M\Vert\equiv\Vert I\otimes A_0+\sum_{i=1}^3\sigma_i\otimes A_i\Vert\geq\Vert A_0\Vert$.

\medskip

With this, we can prove Eq.~\eqref{eq:Bineq}. We first note that $\Vert A\Vert ^2=\Vert A^\dagger A\Vert$ for any operator $A$. Since $\sigma_i$s are Hermitian, we have
\begin{align}
\Bigl\Vert \sum_{i=1}^3\sigma_i\otimes B_i\Bigr\Vert^2&=\Bigl\Vert\sum_{ij}\sigma_i\sigma_j\otimes B_i^\dagger B_j\Bigr\Vert\\
&=\Bigl\Vert I\otimes\sum_{i} B_i^\dagger B_i+\sum_{i\neq j}\sigma_i\sigma_j\otimes B_i^\dagger B_j\Bigr\Vert\nonumber\\
&\geq \Bigl\Vert\sum_i B_i^\dagger B_i\Bigr\Vert\geq \max_i\Vert B_i\Vert ^2,\nonumber
\end{align}
where, in the last line, we have used the above-proven lemma, with $A_0\equiv \sum_i B_i^\dagger B_i$ and recognizing that $\sum_{i\neq j}\sigma_i\sigma_j\otimes B_i^\dagger B_j$ can be written in the form of $\sum_i\sigma_i\otimes A_i$. The final inequality follows from the fact that $\langle\psi| O|\psi\rangle\leq \langle \psi|O|\psi\rangle +\langle \psi|O'|\psi\rangle$, for any $O$ and $O'$ positive semi-definite operators.\qed

\section{Noisy PDD derivation}\label{app:NoisyPDD}
Here, we provide the derivation leading to Eq.~\eqref{eq:NoisyPDD}, for the effective Hamiltonians $K_i$s ($i=0,1,2,3$) for PDD with noisy gates as detailed in the main text. We recall the definitions of the $K_i$s here:
\begin{align}
\upe^{-\upi K_0}&\equiv X\widetilde X \upe^{\upi\tau_\mathrm{P} H},\\
\upe^{-\upi K_1}&\equiv Y\widetilde Z \upe^{\upi\tau_\mathrm{P} H}X,\nonumber\\
\upe^{-\upi K_2}&\equiv Z\widetilde X \upe^{\upi\tau_\mathrm{P} H}Y,\nonumber\\
\upe^{-\upi K_3}&\equiv \widetilde Z \upe^{\upi\tau_\mathrm{P} H}Z,\nonumber
\end{align}
with $\widetilde Z(X)\equiv\upe^{-\upi\tau_\mathrm{P}(H_{Z(X)}+H)}\upe^{-\upi\Gamma_{Z(X)}}$
and $H_{Z(X)}\equiv \frac{\pi}{2\tau_\mathrm{P}} Z(X)$.
The goal here is to work out expressions for $K_i$s, to lowest order in small quantities $\phiSB\delta\equiv \Vert\tau_\mathrm{P}\HSB\Vert$ and $\eta\geq\Vert \Gamma_{X(Z)}\Vert$.

We first need a key formula, which we show here as a lemma,
\begin{equation}\label{eq:keyFormula}
\upe^{A+\gamma B}=\upe^A{\left[1+\gamma{\left(\frac{1-\upe^{-\ad_A}}{\ad_A}\right)}B+O(\gamma^2)\right]},
\end{equation}
where $A$ and $B$ are arbitrary operators, and $\gamma$ is a small scalar parameter.
Here, $\ad_A$ is the map $\ad_A(\cdot)\equiv [A,\,\cdot\,]$, and $\frac{1-\upe^{-\ad_A}}{\ad_A}$ should be understood in terms of a Taylor series,
\begin{equation}
\frac{1-\upe^{-\ad_A}}{\ad_A}=\sum_{n=0}^\infty\frac{(-1)^n}{(n+1)!}(\ad_A)^n.
\end{equation}
\underline{Proof}. We begin with a standard formula available from Lie-algebra textbooks (see, for example, \cite{rossmann2006lie}), applicable to any operator $X$,
\begin{equation}\label{eq:dXdt}
\frac{\upd}{\upd t}\upe^X=\upe^X\frac{1-\upe^{-\ad_A}}{\ad_A}\frac{\upd}{\upd t}X.
\end{equation}
We want to write $\upe^{A+\gamma B}$ as a power series in $\gamma$, 
\begin{equation}
\upe^{A+\gamma B}=\upe^A(1+\gamma C+O(\gamma^2)),
\end{equation}
for some $C$. $C$ can be calculated as
\begin{align}
C&={\left.{\left[\upe^{-A}\frac{\upd}{\upd t}{\left(\upe^{A+\gamma B}\right)}\right]}\right\vert}_{\gamma=0}={\left(\frac{1-\upe^{-\ad_A}}{\ad_A}\right)}B,
\end{align}
where we have made use of the identity Eq.~\eqref{eq:dXdt}. This gives, immediately, formula Eq.~\eqref{eq:keyFormula}, as desired.\qed

Armed with Eq.~\eqref{eq:keyFormula}, we can work out, say, $K_3$. We first observe that $\upe^{-\upi K_3}\equiv \widetilde Z\upe^{\upi\tau_\mathrm{P}H}Z=\widetilde Z Z\upe^{\upi\tau_\mathrm{P}ZHZ}=\widetilde Z Z\upe^{\upi\delta\Omega_3}$, for $\Omega_3\equiv Z\Omega Z=\tau ZHZ$. We compute $\widetilde Z Z$:
\begin{align}
&\quad\widetilde Z Z=Z\upe^{-\upi(\tau_\mathrm{P}H_Z+\delta\Omega_3)}\upe^{-\upi Z\Gamma_Z Z}\\
&\dot{=}{\left\{1\!-\!\upi\tau_\mathrm{P}\!\!\sum_{n=0}^\infty\frac{(-\pi)^n}{(n+1)!}{\left(\!\frac{\ad_Z}{2\upi}\!\right)}^{\!\!n}\!(ZHZ)+O(\gamma^2)\right\}}\upe^{-\upi Z\Gamma_Z Z}\nonumber,
\end{align}
where we have used the key formula Eq.~\eqref{eq:keyFormula} on the exponential $\upe^{-\upi(\tau_\mathrm{P}H_Z+\delta\Omega_3)}$, with $A\equiv -\upi\tau_\mathrm{P}H_Z$ and $\gamma B\equiv -\upi\delta \Omega_3$, so that $\gamma \equiv \delta \phiSB$ is a small parameter. We also note that $\upe^A\dot{=}Z$, where we recall that the notation $\dot{=}$ denotes equality up to an overall phase.

Now, recall that $H=\sum_{\alpha=0}^3\sigma_\alpha\otimes B_\alpha$. We let $h_Z\equiv -X\otimes B_X-Y\otimes B_Y$ so that $ZHZ=h_Z+I\otimes B_I+Z\otimes B_Z$. Observe that $(\ad_Z)^0(ZHZ)=(\ad_Z)^0(h_Z)+I\otimes B_I+Z\otimes B_Z$, and for integer $n>0$, $(\ad_Z)^n(ZHZ)=(\ad_Z)^n(h_Z)$. It is easy to show that, for $n=0,1,2,\ldots$,
\begin{align}
{\left(\frac{\ad_Z}{2\upi}\right)}^n(h_Z)&={\left\{\begin{array}{ll}
(-1)^{n/2}h_Z,&n\textrm{ even},\\
(-1)^{(n-1)/2}h_Z',&n\textrm{ odd},
\end{array}\right.}
\end{align}
for $h_Z'\equiv X\otimes B_Y-Y\otimes B_X$. 
From this, straightforward algebra gives
\begin{align}
&~\quad\upe^{-\upi K_3}=\widetilde Z Z\upe^{\upi\delta\Omega_3}\\
&={\left\{1-\upi\tau_\mathrm{P}(I\otimes B_I+Z\otimes B_Z)+\upi\tau_\mathrm{P}\tfrac{2}{\pi}h_z'+O(\gamma^2)\right\}}\nonumber\\
&\qquad\times\upe^{-\upi Z\Gamma_Z Z}\upe^{\upi\delta\Omega_3}\nonumber\\
&=\exp{\left[-\upi\tau_\mathrm{P}(I\otimes B_I-\tfrac{2}{\pi}X\otimes B_Y+\tfrac{2}{\pi} Y\otimes B_X+Z\otimes B_Z)\right.}\nonumber\\
&\qquad\quad\qquad {\left.-\upi Z\Gamma_ZZ+\upi\delta\Omega_3+O(\gamma^2,\gamma\eta)\right]}\nonumber.
\end{align}
From this, we obtain $K_3$ as given in Eq.~\eqref{eq:NoisyPDD} after further simplification. $K_0,K_1$, and $K_2$ are derived in a similar manner.

\section{Ideal CDD dynamics}\label{app:CDD}
Here, we analyze the dynamics of the system under ideal CDD pulses, giving the derivation to Eq.~\eqref{eq:cdd-generator-est} and related statements used in the main text.

The theory of Lie groups guarantees the existence of a homeomorphism between the Lie-group elements around $I$ and its Lie-algebra elements in the vicinity of $0$. This suggests a one-to-one correspondence between the unitary dynamics $U_n$ for $\CDDn$ and the Hermitian generator $\Omega_{\CDDn} \equiv\upi\log(U_n)$ when noise is weak.
It is more convenient to focus on the generators and regard DD as transformation among them as $n$ changes. 
At the base level, we have the bare Hamiltonian $\Omega\equiv\tau H$.
For $n=1$, CDD gives $\Omega_{\CDD1}\equiv \Omega_{\mathsf{PDD}}$, which can be expressed as a series
 $\sum_{m=1}^\infty \Omega_{\CDD1}\up{m}$ through the Magnus expansion, as we have done in the main text. 
 To characterize this process, we introduce the decoupling maps $\cD$ and $\cD\up{m}$, defined as
 \begin{equation}
  \Omega_{\CDD1} = \cD(\Omega), \quad 
  \Omega_{\CDD1}\up{m} = \cD\up{m}(\Omega),
 \end{equation}
with $\cD \equiv \sum_{m=1}^\infty \cD\up{m}$ following the Magnus series. In particular, $\cD\up{1}$and $\cD\up{2}$ are explicitly given in Eqs.~\eqref{eq:PDDMag1} and \eqref{eq:PDDMag2}. 
Higher-level concatenations are defined iteratively,
\begin{equation}\label{eq:CDD-update}
\Omega_{\CDD{k+1}} = 
\cD(\Omega_{\CDD{k}}) 
= \sum_{m=1}^\infty \cD\up{m}( \Omega_{\CDD{k}} ).
\end{equation}
Backtracking from level $n$ to level $0$, we have
$\Omega_{\CDDn} = (\cD)^{n}(\Omega)$. In the end, 
$\Omega_{\CDDn}$ can be expanded as a Magnus series in the bare Hamiltonian. 
\begin{equation}\label{eq:CDD-magnus}
\Omega_{\CDDn} =\Bigl(\sum_{m=1}^\infty \cD\up{m}\Bigr)^{n}(\Omega)=\sum_{m=1}^\infty \Omega_{\CDDn}\up{m},
\end{equation}
where $\Omega_{\CDDn}\up{m}  \sim  \norm{\Omega}^m$ is the $m$th-order term in $\Omega$.

There is no analytical solution to the full $\Omega_{\CDDn}$. But to estimate CDD dynamics, it suffices to find an analytically solvable estimator $\Ohat_{\CDDn}$ that reflects the leading-order behavior of the full series. We need the leading-order behavior of the pure-bath part and system-bath coupling separately, as the two pieces can be of different orders in small quantities, and we make use of them separately in our analyses. Formally, we define an estimator-error pair,
\begin{equation}
 \Omega_{\CDDn} = \Ohat_{\CDDn} + \delta\Omega_{\CDDn}.
\end{equation}
For an accurate estimator, the error term $ \delta\Omega_{\CDDn}$ must be of higher-order smallness than the estimator, both in the pure-bath part and in the coupling. To quantify this statement, we introduce a two-component norm for any operator $O$ with a split into the bath-only and system-bath coupling terms: $\opnorm{O}\equiv (\Vert O_\rB\Vert,\Vert O_\rSB\Vert)$. Here, we are interested in the two-component norm of both 
$\delta\Omega_{\CDDn}$ and $\Ohat_{\CDDn}$. Accuracy of our estimation scheme demands
\begin{equation}\label{eq:tcn-esterr}
\opnorm{\delta\Omega_{\CDDn}}\!\equiv\!(\delta\phi_{\rB,n},\delta\phi_{\rSB,n})\!\ll \!
\opnorm{\Ohat_{\CDDn}}\!\equiv\!(\widehat\phi_{\rB,n},\widehat\phi_{\rSB,n}),
\end{equation}
for every $n$. Here, $\widehat\phi_{\rB(\rSB),n}$ and $\delta\phi_{\rB(\rSB),n}$ denote the norms of the pure-bath (system-bath coupling) parts of $\Ohat_{\CDDn}$ and $\delta\Omega_{\CDDn}$, respectively. The comparison is implied for both components separately, and should be understood as a comparison of the leading powers of the polynomials, i.e., a smaller polynomial has a larger leading power.

Reference \cite{khodjasteh2007performance} constructed an estimator by keeping the first two Magnus terms for each iteration step. In
our notation, we write this as
\begin{equation}\label{eq:estimator-trunc}
    \Ohat_{\CDD{k+1}} \equiv  \big(
    \cD\up{1} + 
    \cD\up{2}\big) (\Ohat_{\CDD{k}}).
\end{equation}
We explicitly decompose $\Ohat_{\CDD{k}}$ as 
$\sum_i \sigma_i\otimes\wt B_{k,i}$. \Eqref{eq:PDDMag2} suggests that $\wt B_{k,Z}$ vanishes beyond the first level of concatenation. This gives a simple update rule that connects higher concatenation levels,
\begin{align}
 \tau B_{n+1,I} &= 4  \tau B_{n,I} \\
 \tau B_{n+1,X} & = -4\upi [ \tau B_{n,I}, \tau B_{n,X} ],\nonumber\\
 \tau B_{n+1,Y} & = -2\upi  [ \tau B_{n,I}, \tau B_{n,X} ],\nonumber\\
 \tau  B_{n+1,Z} &=0.\nonumber
\end{align}
After a little bit of algebra, the estimator can be shown to be
\begin{align}\label{eq:cdd-estimator}
\Ohat_{\CDDn} &=  I \otimes 4^n \tau B_I \nonumber \\
&+ X \! \otimes\! (-\upi)^{n} 2^{n(n+1)} \tau^{n+1}\!\ad_{B_I}^{n}\!(B_X)\\ 
&+ Y \! \otimes\! (-\upi)^{n} 2^{n^2}\!\!\tau^{n+1}\ad_{B_I}^{n-\!1}\!( [B_I,B_Y] - \upi \{B_X,B_Z\} ).\nonumber 
\end{align} 
To track orders, we focus on the leading powers in $\tau$, ignoring any coefficients.
According to Eq.~\eqref{eq:cdd-estimator}, we have,
\begin{equation}
(\widehat\phi_{\rB,n},\widehat\phi_{\rSB,n} )
\simeq
(\tau,\tau^{n+1}).
\end{equation}
To show that $\Ohat_{\CDDn}$ is indeed faithful, $\delta\phi_{\rB,n}$ and $\delta\phi_{\rSB,n}$ need to be of higher-order smallness compared to $\widehat\phi_{\rB,n}$ and $\widehat\phi_{\rSB,n}$. Indeed, we claim that
\begin{equation}\label{eq:cdd-error-bounds}
(\delta\phi_{\rB,n},\delta\phi_{\rSB,n})\lesssim(\tau^3,\tau^{n+2}).
\end{equation}
We prove this assertion by mathematical induction.

\smallskip
\noindent\underline{Proof}.
For $n=1$, the estimation error comes solely from the series truncation error, which is led by the third-order term,
\begin{equation}
(\delta\phi_{\rB,1},\delta\phi_{\rSB,1})
=\Btrinorm{\sum_{k=3}^\infty \Omega_{\PDD}\up{k}}\Btrinorm \simeq(\tau^3,\tau^3).
\end{equation}
At higher truncation levels, the estimation errors are also supplemented with contributions from the lower-level terms. Formally,
\begin{align}\label{eq:error-update}
&\delta\Omega_{\CDD{n+1}} 
=\cD(\Omega_{\CDDn})-(\cD\up{1}+\cD\up{2})(\Ohat_{\CDDn})\\
&= \cD\up{1}(\delta\Omega_{\CDDn})+ \sum_{m=2}^\infty \cD\up{m}(\Omega_{\CDDn})-\cD\up{2}(\Ohat_{\CDDn}),\notag
\end{align}
where we have used the fact that $\cD\up{1}$ is a linear map. 
To properly bound the size of the error term, we take the two-component norm on both sides of \Eqref{eq:error-update} and apply the triangle inequality:
\begin{align}\label{eq:cdd-error-3terms}
&(\delta\phi_{\rB,n+1},\delta\phi_{\rSB,n+1}) \, \lesssim \, \opnorm{ \cD\up{1}(\delta\Omega_{\CDDn}) }  \; + \; \\
&\, \opnorm{ \cD\up{2}(\Omega_{\CDDn})-\cD\up{2}(\Ohat_{\CDDn})} 
+ \Btrinorm\sum_{m=3}^\infty \cD\up{m}( \Omega_{\CDDn})\Btrinorm .\notag
\end{align}   
We need to show that the right-hand side is bounded by $(\tau^3,\tau^{n+3})$.

Reference \cite{khodjasteh2007performance} demonstrated that the higher-order Magnus series $\sum_{m=3}^\infty\cD\up{m}(\Ohat_{\CDDn})$ is small.
However, Eq.~\eqref{eq:error-update} suggests that this argument is not sufficient, as
the estimation error $\delta\Omega_{n+1}$ comes not only from the truncation error at the same level, but also propagates up from the lower-level error $\delta\Omega_{\CDDn}$. 
Let us examine the size of these terms. The first term in (\ref{eq:cdd-error-3terms}) is straightforward,
\begin{equation}
\opnorm{\cD\up{1}(\delta\Omega_{\CDDn})} =
(4 \delta\phi_{\rB,n},0) \simeq
(\tau^3 ,0).
\end{equation}
To bound the second term in (\ref{eq:cdd-error-3terms}), we need to calculate the difference between the full $\Omega_{\CDDn}$ and $\Ohat_{\CDDn}$ after applying $\cD\up{2}$.
Using the explicit expression for the second-order Magnus term, we obtain
\begin{align}
&\quad~ \opnorm{\cD\up{2}(\Omega_{\CDDn})-\cD\up{2}(\Ohat_{\CDDn})}\\
&\simeq  (0, \delta\phi_{\rB,n} \phi_{\rSB,n} +\phi_{\rB,n} \delta\phi_{\rSB,n})\simeq (0,\tau^{n+3}),\nonumber
\end{align}
where we have applied the induction hypothesis $\phi_{\rB,n}\simeq\widehat\phi_{\rB,n}\simeq\tau$ and $\phi_{\rSB,n}\simeq\widehat\phi_{\rSB,n}\simeq \tau^{n+1}$. The final term in (\ref{eq:cdd-error-3terms}) involves an infinite sum of Magnus terms higher than the third order. Since the series is led by the third-order term, which can be explicitly bounded, we have
\begin{align}
&\quad \Btrinorm\sum_{m=3}^\infty \cD\up{m}( \Omega_{\CDDn})\Btrinorm \simeq 
 \opnorm{\cD\up{3}(\Omega_{\CDDn})} \\
& \qquad\quad \lesssim (\phi_{\rSB,n}^2 (\phi_{\rB,n}+\phi_{\rSB,n}) ,\phi_{\rSB,n} (\phi_{\rB,n}+\phi_{\rSB,n})^2)\nonumber\\
&\qquad\quad \simeq (\tau^{2n+2},\tau^{n+3}).\nonumber
\end{align}
The estimation for $\cD\up{3}(\Omega_{\CDDn})$ can be done by explicitly calculating the third-order Magnus term for PDD, whose expression is not relevant here.
Since all three terms are bounded by $(\tau^3, \tau^{n+3})$, so is their sum. This proves our assertion Eq.~\eqref{eq:cdd-error-bounds}.
\qed

\smallskip

We have now shown that the estimator constructed through Eq.~\eqref{eq:estimator-trunc} is indeed accurate.
This leads to the conclusion that ideal $\CDDn$ indeed achieves $n$th-order decoupling, with leading-order behavior given by $\Ohat_n$ in Eq.~\eqref{eq:cdd-estimator}, which is also Eq.~\eqref{eq:cdd-generator-est} in the main text.

\end{document}